\documentclass[aps,prb,twocolumn,superscriptaddress]{revtex4-2}

\usepackage{amsmath , amssymb,  graphicx}%, subcaption}

\usepackage[colorlinks=true ,urlcolor=blue,urlbordercolor={0 1 1}]{hyperref}

\usepackage{color}
\usepackage{xcolor}
\usepackage{braket}

\usepackage[utf8]{inputenc}
\usepackage{bm}
\usepackage{booktabs}
\usepackage{verbatim}
\usepackage{graphicx}
\usepackage{amsmath}
\usepackage{color}
\usepackage{float}
\usepackage{bbm}
\usepackage{amssymb}
\usepackage{slashed}
\usepackage{wasysym,bm,bbm,dsfont}
\usepackage{makecell}
\usepackage[normalem]{ulem}

\begin{document}

\title{Resonating-valence-bond superconductor from small Fermi surface in twisted bilayer graphene}

\author{Jing-Yu Zhao}
\email{zhaojingyu15@gmail.com}
\affiliation{Department of Physics and Astronomy, Johns Hopkins University, Baltimore, Maryland 21218, USA}

\author{Ya-Hui Zhang}
\email{yzhan566@jhu.edu}
\affiliation{Department of Physics and Astronomy, Johns Hopkins University, Baltimore, Maryland 21218, USA}

\date{\today}

\begin{abstract}
%Mechanism of superconductivity in twisted bilayer graphene (TBG) remains one of the central problems in correlated moir\'e materials. The most intriguing question is about the nature of the normal state: is the Cooper pair formed from small Fermi surface or large Fermi surface?  In this work we point out the possibility of a symmetric pseudogap metal with small hole pockets, dubbed as second Fermi liquid (sFL).   In the sFL phase at $\nu=-2-x$, there is a two-component picture: two electrons mainly localize at each AA site and form a paired singlet due to anti-Hund's coupling mediated by the optical phonon, while additional holes doped into the AA sites form small Fermi surfaces. The sFL phase corresponds to an intrinsically strongly interacting fixed point and is topologically distinct to the conventional Fermi liquid. We develop a unified framework to describe both a renormalized  FL phase and an sFL phase.  We propose that the  TBG superconductor emerges from the sFL phase, but is close to the transition toward the FL phase under increasing hole doping. In the superconductor, pairing of local moments is shared  to the mobile carriers and a smaller superconducting gap with nodal $p_x$ symmetry is opened on the small hole pockets.  This work provides, to our knowledge, the first unified theory that explains both the pseudogap metal above $T_c$ and the two-gap nematic superconductivity below it. 
Understanding the mechanism of superconductivity in twisted bilayer graphene (TBG) is a central challenge in correlated moiré materials. 
The most intriguing question is about the nature of the normal state: is the Cooper pair formed from small Fermi surface or large Fermi surface?
Here, we propose a unified theoretical framework describing a symmetric pseudogap metal with small hole pockets, dubbed the second Fermi liquid (sFL), which is topologically distinct from the conventional Fermi liquid. 
Our theory shows that superconductivity in TBG emerges from the sFL phase, with local moment pairing transferred to mobile carriers, giving rise to a smaller nodal superconducting gap on the small hole pockets. 
This framework simultaneously explains the pseudogap metal above the transition temperature and the two-gap nematic superconductivity below it, providing a comprehensive understanding of correlated physics in TBG.
\end{abstract}

\maketitle

\section*{Introduction}

The nature of superconductivity in magic-angle twisted bilayer graphene (TBG) \cite{cao2018unconventional,Cao2018mott,Yankowitz2019layer,Lu2019moreSC,Stepanov2020,Cao2021nematic,Liu2021Coulomb,arora2020superconductivity} and in twisted multilayer graphene \cite{park2021tunable,hao2021electric,park2022robust,cao2021pauli} remains a central puzzle in the study of correlated moir\'e materials \cite{andrei2020graphene,andrei2021marvels,nuckolls2024microscopic}. Various pairing mechanisms have been proposed, including electron-phonon coupling \cite{wu_ephsc_2018,lian2019twisted,chou2019superconductor,wu2019phonon}, weak-coupling theories \cite{sharma_plasma_2020, isobe_vanHove_2018,chichinadze_vanHove_2020, kennes_frg_2018, gonzalez_kohn_luttinger_2019,you2019superconductivity, huang_pseudospin_2022,wang_kekule_2025}, and skyrmion-mediated superconductivity \cite{khalaf2021charged}. However, a comprehensive and well-established theory remains elusive. In particular, recent tunneling spectroscopy measurements \cite{park_pseudogap_2025,kim_twogap_2025} have observed two distinct superconductivity like gaps in the superconducting phase. 
The smaller gap closes at the critical temperature $T_c$, while the larger gap survives above $T_c$, persisting as a pseudogap in the normal state.  To our knowledge, no existing theory provides a unified explanation for both this two-gap structure and the associated pseudogap.

The existence of the pseudogap phase is reminiscent of high-$T_c$ cuprates \cite{lee2006doping}.  This naturally raises the question of whether superconductivity in TBG should also be understood as arising from doping a Mott insulator. However, applying a conventional Hubbard model to TBG is precluded by its fragile band topology \cite{Po2018IVC,Tarnopolsky2019,Ahn2019fgtop,Ledwith2021Peda,Song2021symano}. As a result, many theoretical studies have focused on symmetry-breaking phases, such as inter-valley coherent (IVC) orders \cite{bultinck2020ground,kwan2021kekule,parker2021strain,wagner2022global,wang_IKS_2025}, often identified using momentum-space Hartree-Fock calculations within the continuum model \cite{Bistritzer2011}. Indeed, an intervalley Kekulé spiral (IKS) state -- a specific IVC order with a non-zero wavevector $\mathbf Q$ \cite{wagner2022global, kwan_kekule_2021} -- has been observed at filling $\nu=-2$ in STM experiments \cite{kim2023imaging, nuckolls2023quantum,kim_twogap_2025}. However, the relationship between this IKS state and superconductivity remains unclear. The lesson from cuprates suggests that the superconducting phase at finite hole doping may be disconnected from the symmetry-breaking order of the parent state at zero doping. 
Moreover, pseudogap physics may already develop at temperatures far above the onset of the ordered state. 
Therefore, it remains worthwhile to explore theories of normal Fermi liquid states before IVC orders develop, as well as superconductivity emerging directly from these parent states.

Recently, the relevance of Mott physics and the formation of local moments\cite{rozen2021entropic,saito2021isospin} in TBG have been increasingly recognized.  
Due to the fragile topology of the active bands, such Mott physics cannot be described by a conventional local-orbital Hubbard model, and has instead been studied using lattice models such as the topological heavy fermion model (THFM) \cite{Song2022THFM,Calugaru2023THFM,Yu2023THFM,Hu2023THFM,Vafek2024THFM,Zhou2024THFM,Hu2023THFM2,Chou2023THFM,Wang2024THFM,Lau2023THFM,Rai2024THFM,Youn2024DMFT} and model with non-local Wannier orbital \cite{Ledwith2024}. Alternatively, Mott physics can be understood using the ancilla framework \cite{zhao_tbg_2025} proposed by one of us \cite{Zhang2020}. Within this framework, it was demonstrated that a symmetric Mott state is possible at $\nu=-2$. Upon hole doping, the quasi-particles are still mainly from the $f$ orbital instead of the $c$ orbital in THFM \cite{zhao_tbg_2025,zhao_mixed_2025,ledwith2025exotic}. Specifically, at filling $\nu=-2-x$, this approach predicts a symmetric pseudogap metal with small hole pockets.

In this work, we propose that the normal state of the TBG superconductor is exactly the pseudogap metal discussed above, which we dub the second Fermi liquid (sFL). The unique symmetry of TBG, $(U(1)_{K}\times U(1)_{K'}\times SU(2))/Z_2$, permits two distinct, symmetric Fermi liquid fixed points \cite{zhang2020spin,yang2024strong}: a conventional Fermi liquid (FL) with a large Fermi surface, and an sFL phase with a small Fermi surface.  
Both states can in principle serve as parent normal states for further low-temperature instabilities. 
In this work, we focus on the superconducting instability emerging from the sFL parent state. 
This focus does not exclude possible symmetry-breaking tendencies, such as IVC order, 
which may become important near the correlated insulator at $\nu=-2$.  
However, a systematic study of such competing orders is beyond the scope of this work. 
%Near filling $\nu=-2-x$, the relevant normal state is dominated by the sFL. }
The sFL phase is characterized by an emergent two-component picture: two electrons form localized moments on the AA sites, which are then paired into singlets by an on-site anti-Hund's coupling $J>0$ mediated by optical phonons \cite{chen2024strong}. The additional carriers, with density $x$, form small hole pockets.  We emphasize that this two-component picture is emergent and should not be confused with the naive decoupling of itinerant $c$ and localized $f$ orbitals in the THFM. Instead, both the mobile carriers and the local moments are from the $f$ orbital.  Conceptually, our sFL phase is closer to the fractionalized Fermi liquid (FL*) in one-orbital model of cuprate\cite{Zhang2020} than a naive Kondo decoupled phase, but sFL phase does not have deconfined spin-liquid or fractionalization and is essentially still a Fermi liquid. 

The pairing of local moments, with an energy scale $J$, is consistent with experimental observations of a pseudogap, which manifests itself  as broad peaks in $dI/dV$ at $\omega = \pm \Delta_{\mathrm{PG}}$ \cite{park_pseudogap_2025,kim_twogap_2025}.  Due to strong repulsion $U$, these localized pairs are immobile and thus cannot directly lead to superconductivity. A key contribution of this work is to demonstrate that a superconducting dome emerges in proximity to the transition from the sFL to a renormalized FL phase, tuned by decreasing $J$ or increasing doping $x$. We develop a parton mean-field theory based on the THFM to provide a unified framework for both the sFL and FL phases. Within this theory, approaching the FL phase from the sFL side triggers the condensation of a slave boson $B$ below a coherence temperature $T_{\mathrm{coh}}$. Below $T_{\mathrm{coh}}$, this condensation allows the pre-formed pairing of the local moments to be transferred to the mobile carriers, opening a small superconducting gap on the small Fermi surfaces.  The secondary superconducting pairing is characterized by a large real-space size and a nodal $p_x$ pairing symmetry around the Fermi surface.

The spirit of our theory is similar to the Resonating Valence Bond (RVB) theory proposed for cuprates \cite{anderson1987resonating}.  However, our model differs in key aspects: (I) Our parent state is a valence bond state without fractionalization, not an RVB spin liquid. (II) While both theories assume pre-formed pairs, their role differs. In conventional RVB theory\cite{lee2006doping}, the spinon pairing in the parent state evolves directly into the superconducting pairing, while in our theory a secondary pairing is induced for the additional mobile carriers. Our mechanism also has conceptual similarities to the ``molecular pairing'' scenario in Ref.~\cite{wang2024molecular}, but that analysis was restricted to a single Kondo impurity model. Also the mobile carrier in our theory is also from the $f$ orbital, and the physics  aligns more closely with a purely $f$-orbital Hubbard model\cite{zhang2020spin} rather than with  Kondo model analysis presented in Ref.~\cite{wang2024molecular}.

\begin{figure}[t]
    \centering
    \includegraphics[width=0.98\linewidth]{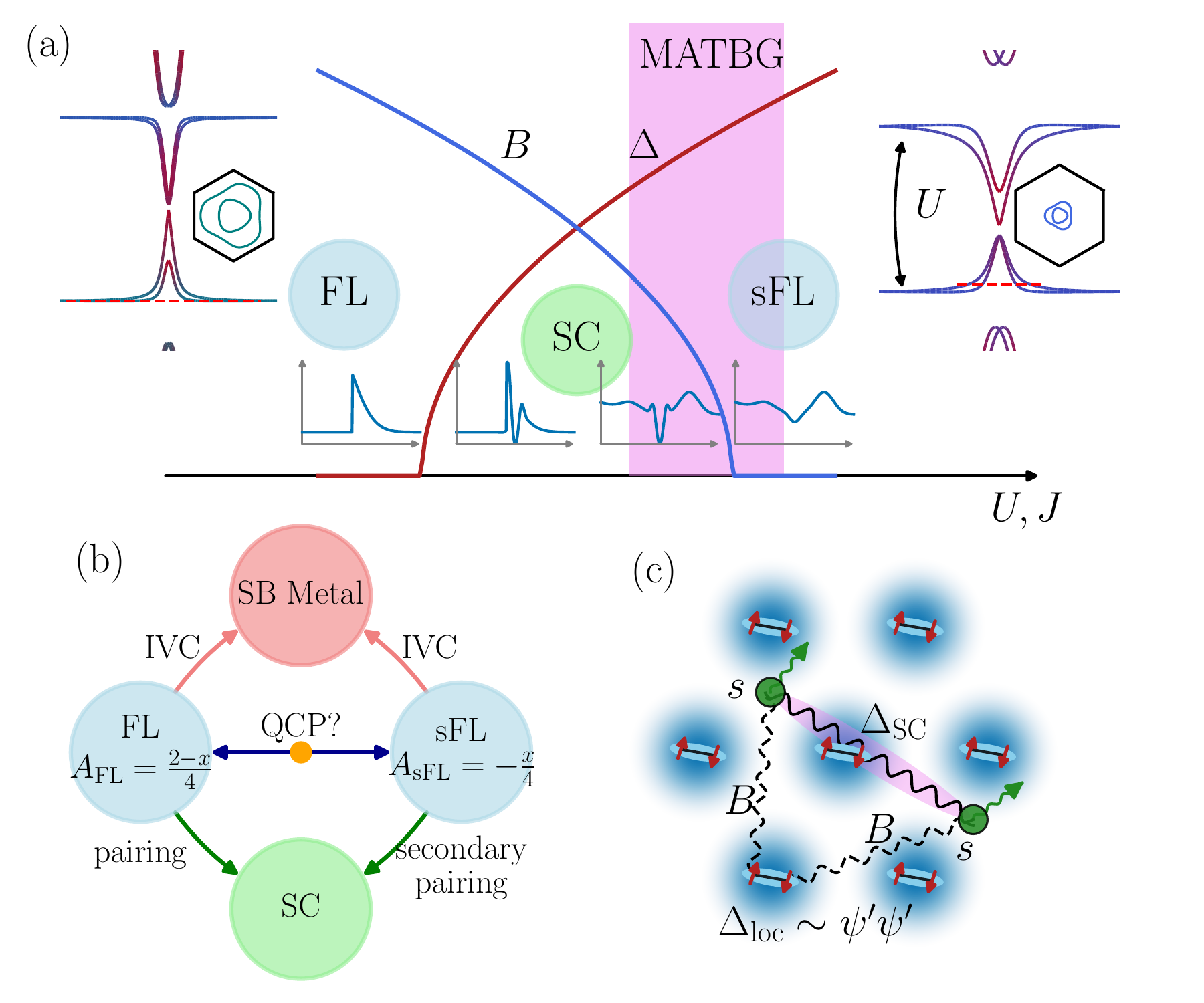}
    \caption{
    \textbf{Schematic diagrams of the sFL, FL and superconducting phases.}
    (a) Schematic phase diagram illustrating the evolution from the sFL to the FL phase, for example, tuned by the anti-Hund's coupling $J$. The vertical axis denotes the temperature $T$. The left and right insets show the band structures and Fermi surfaces of the renormalized FL and sFL phases, respectively. $\Delta$ is the pairing of the local moments from the $J$ term. $B$ is the slave boson condensation, which sets a coherence temperature scale $T_{\mathrm{coh}}$. In the phase diagram we also show illustration of tunneling spectrum.
    (b) Schematic illustration of the relationships among different phases. The FL and sFL phases correspond to two distinct symmetric fixed points.
    Both Fermi liquids can serve as parent states of symmetry breaking metals or superconductivity. 
    (c) Illustration of the RVB mechanism of superconductivity. We already have local pairing of spinons $\Delta_{\mathrm{loc}}$. Onset of the slave boson condensation $B$ then induces resonance between the local pairing and two mobile carriers, leading to a superconducting pairing $\Delta_{\mathrm{SC}}$ between the mobile carriers, which can be well separated in space.
    }
    \label{fig:illu}
\end{figure}

\begin{figure}[t]
    \centering
    \includegraphics[width=0.98\linewidth]{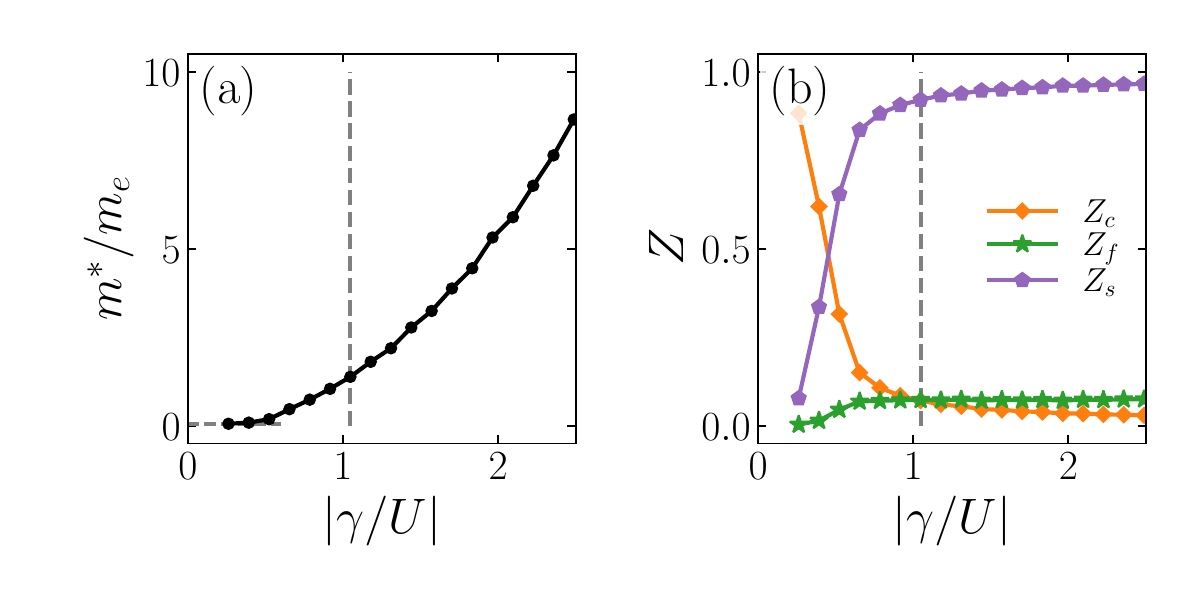}
    \caption{ 
    \textbf{Effective mass and quasiparticle weights across the sFL regime.}
    (a) The effective mass $m^*/m_e$ as a function of $\gamma/U$ for the sFL phase at $\nu=-2.4$, where effective mass $m^*$ is estimated by the zero energy density of state and $m_e$ is the free electron mass. 
    The horizontal dashed line marks the free electron mass of the $c$ electron in the decoupled limit, with $m^*_0/m_e=0.05$.  
    (b) the quasi-particle weight of $c$, $f$ and $s$ fermions as a function of $\gamma/U$ for $\nu=-2.4$. 
    All the calculations are performed by varying $U$ at fixed $\gamma = -26.184$ meV, with $w_0/w_1=0.8, \theta=1.06^\circ$ and $J=6.0$ meV. 
    The value of $\gamma/U$ used in the main text is indicated by the vertical dashed lines in panels (a) and (b). 
    The slave boson $B$ is artificially set to 0 in the mean-field iteration to get the normal state of sFL. 
    }
    \label{fig:mass}
\end{figure}

\section*{Results}

\textbf{Model} We use the THFM  
\cite{Song2022THFM,Calugaru2023THFM,Yu2023THFM, Hu2023THFM, Vafek2024THFM, Zhou2024THFM, Hu2023THFM2, Wang2024THFM, Lau2023THFM, Rai2024THFM,Youn2024DMFT, Hu2025THFM} 
description of the low energy bands of TBG: 
\begin{equation}\label{eqn:THFM_0}
    H= H^{(c_1,c_2)}_0  + H_0^{(c_1, f)} + H_{\mathrm{int}}^{(f)}-\mu (N_c+N_f)~,
\end{equation}
which includes two dispersive bands, $c_1$ and $c_2$ described by $H_0^{(c_1,c_2)}$, and a localized orbital $f$ on AA site.
The $c$ and $f$ bands are hybridized through $H_0^{(c_1,f)}$.   
We use $f_{i}$ and $c_{1,\mathbf{k}}$, $c_{2,\mathbf{k}}$  as eight-component spinors, collecting spin, valley, and orbital flavor: $f_{i} = \{f_{i;\alpha}\}$, $c_{1,\mathbf{k}} = \{c_{1,\mathbf{k};\alpha}\}$ and $c_{2,\mathbf{k}} = \{c_{2,\mathbf{k};\alpha}\}$, with $\alpha=a \tau s$ formed by the orbital $a=\pm$, valley $\tau=K, K'$ and spin $s=\uparrow, \downarrow$. 
Here we choose the orbital index $a=\pm$ such that $f_{a=\pm;\tau s}$ has angular momentum $L=\pm 1$ around each AA site.
%where both $f_{\mathbf{k}}$ and $c_{1,\mathbf{k}}$ are eight-component spinors, collecting spin, valley, and orbital flavor: $f_{\mathbf{k}} = \{f_{\mathbf{k};\alpha}\}$ and $c_{1,\mathbf{k}} = \{c_{1,\mathbf{k};\alpha}\}$, with $\alpha = a \tau s$ formed by the orbital $a=\pm$, valley $\tau=K, K'$ and spin $s=\uparrow, \downarrow$. 
%We choose $a$ such that $f_{a=\pm;\tau s}$ has angular momentum $L=\pm 1$ around each AA site.
For the $f$ orbital,  we include on-site interaction:
\begin{equation}\label{eqn:interactionf}
    H_{\mathrm{int}}^{(f)} = \frac{U}{2}\sum_i(n_{i;f} - 4-\kappa\nu)^2 + \sum_ih_{i;J}^{(f)}, 
\end{equation}
where we follow Ref.~\cite{Lau2023THFM} to add a phenomenological parameter $-\kappa \nu$ in the Hubbard interaction.  
We also include an intra-site spin interaction term $h_{i;J}^{(f)}$ arising from electron–electron and electron-phonon couplings~\cite{Wang2024THFM}. 
$h_{i;J}^{(f)}$ favors inter-valley spin-singlet pairing of the local moments as discussed in the Supplementary Material. 

%\textbf{Simplified model at $\nu=-2-x$} 
Different valences of the local $f$ orbital are well separated on energy by the Hubbard $U$. We label the states with $n_{i;f}=0, 1,2,3$ at each AA site $i$ as holon,  singlon, doublon and triplon. 
In the vicinity of the $\nu=-2$, we can consider a simplified model by projecting the original Hamiltonian into a restricted Hilbert space including only the 8 singlon state $\ket{i;\alpha}_s=f^\dagger_{i;\alpha}\ket{0}$, 28 doublon states $\ket{i;\alpha \beta}_d=f^\dagger_{i;\alpha} f^\dagger_{i;\beta} \ket{0}$ (with $\alpha<\beta$) and 56 triplon states $\ket{i;\alpha \beta \gamma}_t=f^\dagger_{i;\alpha}f^\dagger_{i;\beta}f^\dagger_{i;\gamma} \ket{0}$ (with $\alpha<\beta<\gamma$) at each AA site $i$. The itinerant $c_1,c_2$ bands are still treated as free fermions.  In the regime of $U\gg J$, we ignore the holon state, thus the pairing mechanism through virtual Cooper pair proposed in Ref.~\cite{yang2024strong,oh2024high} does not apply here.

In the restricted Hilbert space, 
the interaction term reads 
\begin{equation}\label{eqn:THFM_U_esd}
    P_GH_{\mathrm{int}}^{(f)}P_G = \sum_i(E_s n_{i;s} + E_t n_{i;t} + h^{(f)}_{i;J}+\mathrm{const.})~,
\end{equation}
where $E_s = U/2 + U(2+\kappa\nu)$ and $E_t = U/2 - U(2+\kappa\nu)$ are on-site energy of singlon and triplon states. $n_{i;s}$, $n_{i;d}$ and $n_{i;t}$ are the density of singlon, doublon and triplon.

\textbf{Parton Mean field theory} One convenient way to deal with the Hilbert space restriction is to use a parton theory.  We imagine the singlon, doublon and triplon states are created by independent fermionic particles:  $\ket{i;\alpha}_s=s^\dagger _{i;\alpha}|0\rangle$, $\ket{i;\alpha \beta}_d=\psi'^\dagger_{i;\alpha}\psi'^\dagger_{i;\beta} |0\rangle$ and $\ket{i;\alpha \beta \gamma}_t=t^\dagger_{i;\alpha\beta\gamma}|0\rangle$.  

The physical $f$ operator in the restricted Hilbert space can be rewritten as
\begin{equation}\label{eqn:frac}
    f_{i;\alpha} = \sum_\beta s^\dagger_{i;\beta}\psi'_{i;\beta}\psi'_{i;\alpha}
    +\sum_{\beta\gamma(\beta<\gamma)} t^{}_{i;\alpha\beta\gamma} \psi'^\dagger_{i;\beta}\psi'^\dagger_{i;\gamma}~.
\end{equation}
At each site $i$, the partons always satisfy the local constraint
$n_{i;s} + n_{i;\psi'}/2 + n_{i;t} = 1$, under which we have the relation $n_c+n_t-n_s=-x$ on average. 
We can think that  $s,\psi',t$ carry physical  charge $+1,0,-1$, respectively. 
The $\psi'$ fermion is therefore a neutral spinon.

We can subsitute the above parton construction to the Hamiltonian and then do mean field decoupling.  The spin interaction $h_{i;J}^{(f)}$ favors inter-valley singlet state between either $a\eta s$ and $\bar a\bar\eta\bar s$ ($s$-wave) or between $a\eta s$ and $a\bar\eta\bar s$ ($d$-wave).   In the main text we focus on the time reversal invariant $d$-wave ansatz 
which can be stabilized by a small hetrostrain, as discussed in more detail in the Supplementary Material. 
We note that the sFL and pseudogap physics are essentially independent of the pairing channel, and only the low-energy superconducting gap is affected. 
The mean field Hamiltonian reads  $H_{J,\mathrm{MF}} = -2J \Delta^* \sum_{i,\alpha}\alpha\psi'_{i; \bar\alpha} \psi'_{i;\alpha}$, 
where $\bar \alpha = a\bar\eta\bar s$ denotes the pairing partner of $\alpha = a\eta s$, 
and $\alpha=\pm$ labels the corresponding spin $\uparrow,\downarrow$ when it does not appear as a subscript. 

With $\Delta$, we can describe the sFL phase with pairing of spinon $\psi'$. To describe a FL phase, we need to condense a slave boson, a bound state between electron and spinon $B\sim f^\dagger_\alpha \psi'_\alpha$.  Together, we obtain the mean-field form of the $f$ fermion $f_\alpha\sim B\psi'_{\alpha} + \alpha \Delta s^\dagger_{\bar\alpha} + \Delta t_\alpha$. 
The ansatz make it possible to capture FL, sFL phase and superconductor (SC) phase in a unified framework.  The corresponding ansatz are:

\begin{itemize}
    \item sFL: $\langle \Delta \rangle \neq 0$, $\langle B \rangle =0$.    There are hole pockets with Fermi surface volume (per spin and valley) $A_{\mathrm{FS}}=-\frac{x}{4}$, mainly formed  by the singlon $s^\dagger$ when $x$ is relatively large.  
    \item FL: $\langle \Delta \rangle = 0$, $\langle B \rangle \neq 0$. Now $f\sim s  \sim \psi'$. The Fermi surface area per spin-valley flavor is $A_{\mathrm{FS}}=\frac{2-x}{4}$.  
    \item SC: $\langle \Delta\ \rangle \neq 0$, $\langle B \rangle \neq 0$. We can reach the SC phase from either the FL or sFL phase.
\end{itemize}

\begin{figure}
    \centering
    \includegraphics[width=0.95\linewidth]{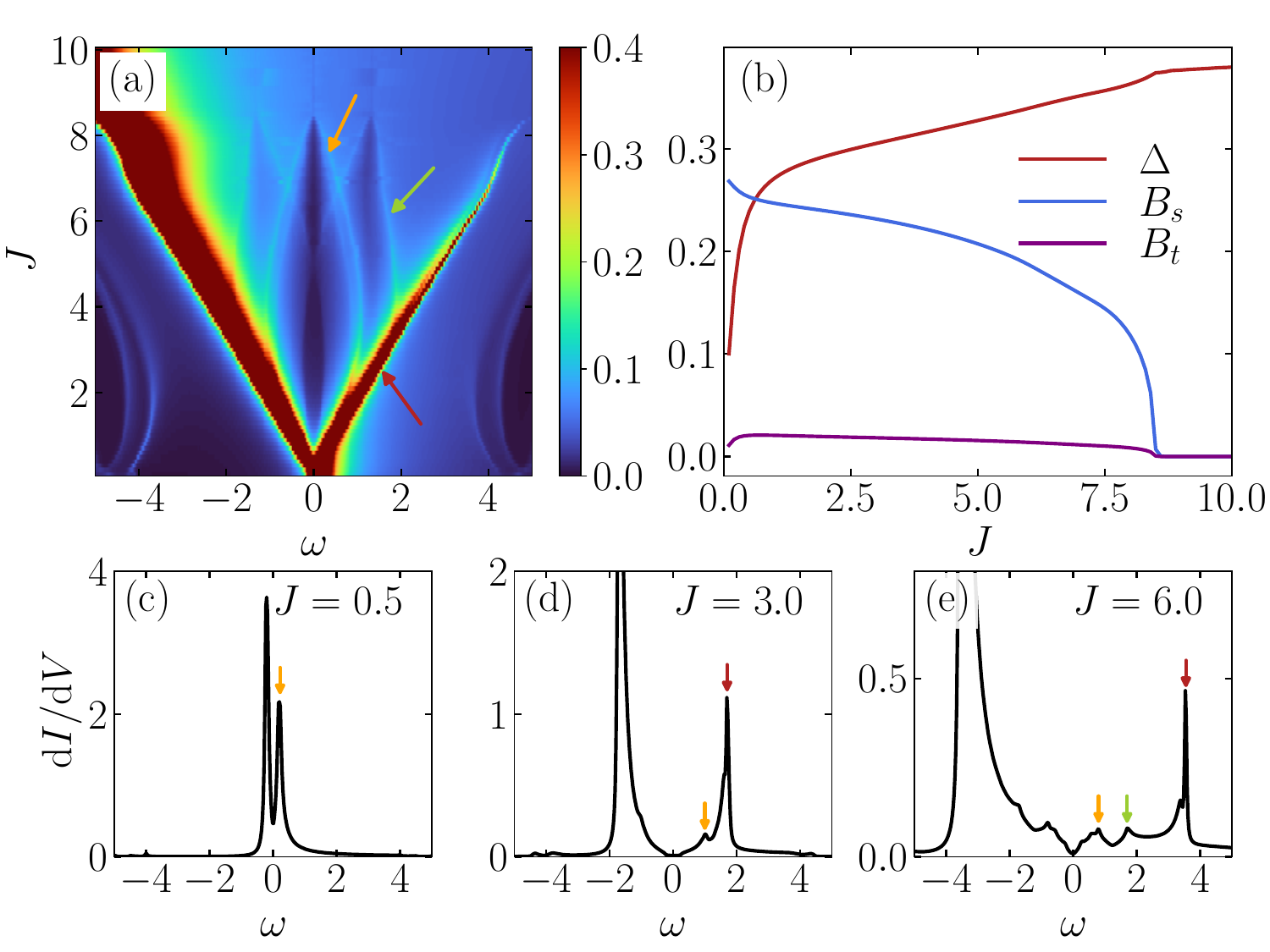}
    \caption{\textbf{Superconducting evolution as a function of the on-site spin interaction $J$, for $\theta=1.06^\circ, w_0/w_1=0.8$, $U=25$ meV, $x=0.4$ and $M=0$. }
    (a) The STM spectrum and (b) the order parameters for different value of $J$. The orange, red and green arrows mark the SC gap, the pseudogap and the inter-band gap. 
    While the pseudogap always increase with increasing $J$, the SC gap first increases and then decreases as a function of $J$. 
    (c) (d) and (e) show three different line cuts of STM at $J=0.5$ meV, $J=3.0$ meV and $J=6.0$ meV, respectively. 
    }
    \label{fig:stm_JA}
\end{figure}

\begin{figure}[t]
    \centering
    \includegraphics[width=0.95\linewidth]{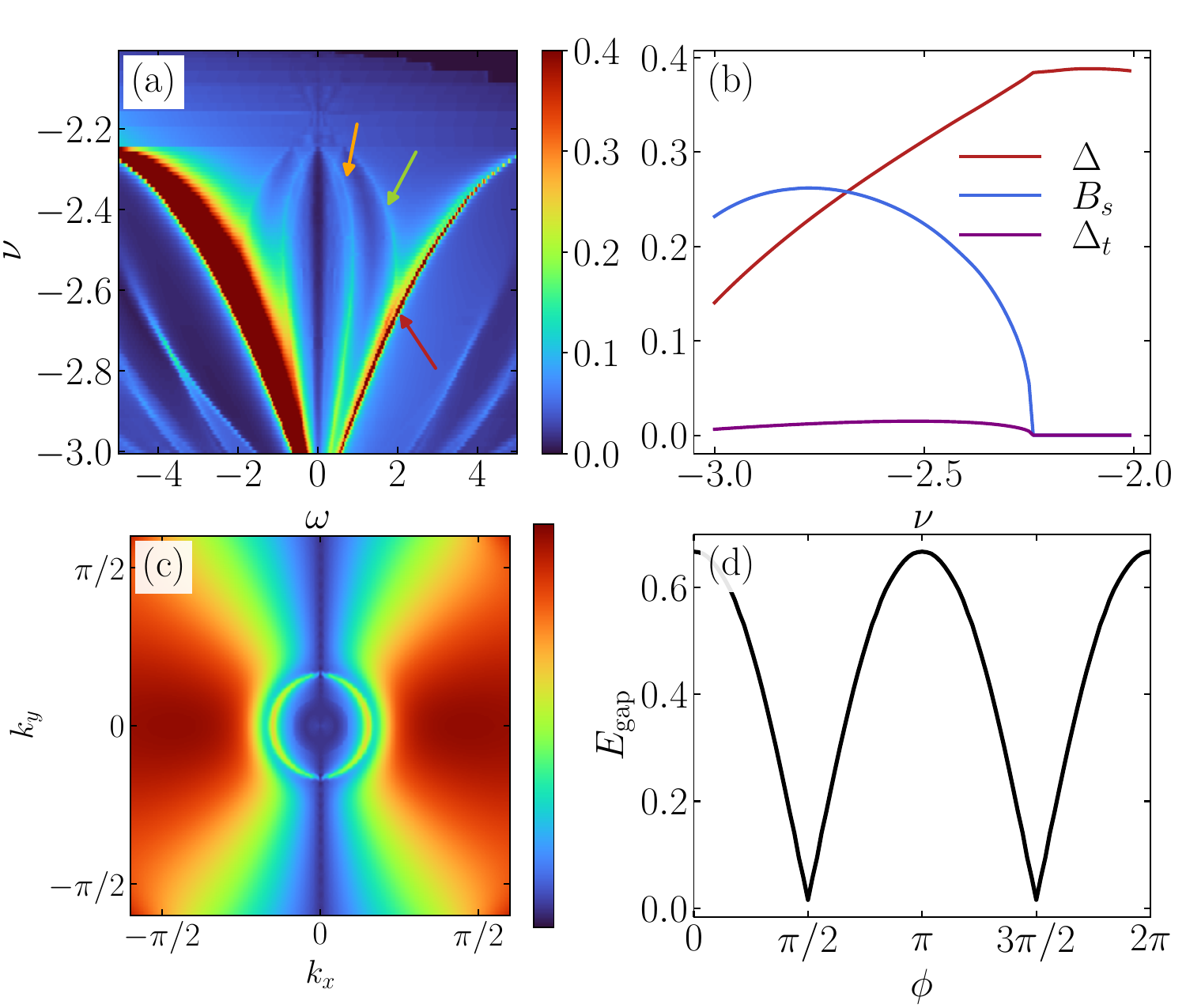}
    \caption{\textbf{Superconducting evolution as a function of the carrier density $\nu$ for twist angle $\theta=1.06^\circ$, $w_0/w_1=0.8$, $U=25$ meV, $J=6.0$ meV and $M=0$. }
    (a) The STM spectrum as for different charge densities. 
    The orange, red and green arrows mark the SC gap, the pseudogap and the inter-band gap. 
    (b) the order parameters for different value of $\nu$.  
    %(c) (d) and (e) show three different line cuts of STM at $\nu=-2.9$, $\nu=-2.6$ and $\nu=-2.3$, respectively. 
    (c) and (d) show the nematic $d$-wave structure at a specific filling $\nu = -2.5$.
    (c) The pairing order parameter $|\Delta_{\mathrm{act,1}}(\mathbf{k})|$, projected onto one of the lower Hubbard bands obtained by artificially setting $B_s = 0$. 
    (d) The minimal single particle gap $E_{\mathrm{gap}}$ along different azimuth direction $\phi$. 
    The gap reaches its maximum along $\phi =0,\pi$ directions and its minimum along $\phi \approx \pm \pi/2$ directions. }
    \label{fig:stm_noproj}
\end{figure}

\textbf{Correlated insulator and sFL} We first consider the ansatz with $\Delta \neq 0$, but $B=0$. At $\nu=-2$, this ansatz describes a mixed-valence Mott insulator \cite{zhao_mixed_2025} with a small gap between the two Hubbard bands. We focus on the lower Hubbard band, the excitation is dominated by a composite fermion 
%$\psi_i= \frac{2}{\sqrt{3}}\{f_{i;\alpha},n_{i}\}- 3\sqrt{3}f_{i;\alpha}$ 
$\psi_{i;\alpha}= -\frac{\sqrt{3}\alpha}{2} s^\dagger_{i;\bar\alpha}+\frac{1}{2}t_{i;\alpha}$ 
\cite{zhao_mixed_2025} at $k=0$. But away from $k=0$, the quasiparticle is dominated by $s^\dagger$.

At $\nu=-2-x$, we have small hole pockets by moving the chemical potential to the lower Hubbard band (see the right inset of Fig.~\ref{fig:illu}(a)). For each spin-valley flavor, there are two separate hole pockets arising from $s_+, s_-$. They are hybridized in the form  $\sim(Me^{2i\phi(\bf k)} 
+ U\frac{v'_\star|\mathbf{k}|}{\gamma} e^{-i\phi(\bf k)})s_{\mathbf{k};+}^\dagger s_{\mathbf{k};-}^{}+\mathrm{h.c.}$, 
where $M$ is the active band width and  $\phi(\mathbf k)$ labels the angle of the momentum $\mathbf k$.
We always have two separate small hole Fermi surfaces. Around each Fermi surface, the spinor in the basis of $(s_+, s_-)$ shows a non-trivial winding. 

We emphasize that the sFL phase described here is beyond the naive decoupling limit between $c$ and $f$ at $\gamma/U=0$. In Fig.~\ref{fig:mass} we show the evolution of the effective mass and the quasi-particle residue versus $\gamma/U$. When $\gamma/U \sim 1$ as used in our calculations relevant to TBG, $m^*$ reaches order one of $m_e$ and is roughly 20 times heavier than the bare itinerant $c$ band. Moreover, the quasi-particle is dominated by the $s^\dagger$ fermion rather than $c$ fermion. 
Clearly, both the mobile carriers and the local moments are mainly from the $f$ orbital, so the essential physics appear to be closer to Hubbard model than Kondo model. In our theory, the  itinerant $c$ bands only  bridge different AA sites and provide an effective dispersion for the $f$ band, but they do not contribute to either Fermi surface or superconductivity directly.

\textbf{Nematic superconductor from sFL: two-gap structure } The sFL phase itself is expected to be stable when $J$ is large.   However, when $J$ is reduced, the spinons can also gain coherence and the system transits to the FL phase.    In Fig.~\ref{fig:stm_JA}, we show the mean-field results by tuning the spin interaction $J$ at a given doping level $x=0.4$.  Indeed, we find an onset of $B$ when decreasing $J$, leading to a superconducting phase with both $B$ and $\Delta$ non-zero. 
Note that when $B$ is finite, the pairing of spinons (energy scale $J\Delta$) can induce a smaller superconducting pairing $\Delta_{\mathrm{SC}}$ for the mobile carriers. Fig.~\ref{fig:stm_JA}(a) shows the tunneling spectrum as a function of $J$.  At large $J$, the SC is from the sFL phase, exhibiting two distinct gaps: a pseudogap $\Delta_{\mathrm{PG}}$ labeled by the red arrow with the energy scale of $J\Delta$, and a superconducting gap $\Delta_{\mathrm{SC}}$ labeled by the orange arrow. The superconducting gap scales with $B^2$ at small $B$.  In the large $J$ regime, we can also see another peak in tunneling spectrum labeled by the green arrow, which is from inter-band pairing.   When decreasing $J$, the pseudogap and the SC gap merge together. At very small $J$, the SC should be understood as from the FL phase, then there is only one superconductor gap decided by $J \Delta$.  Note the FL phase has a Kondo-like peak, and the superconductor gap simply splits it. An illustration of the band and Fermi surface of the FL phase can be found in Fig.~\ref{fig:illu}(a).

To match the experimental phenomenology, we believe TBG is in the sFL side, so a relatively large $J$ is required. We then fix $J=6$ meV and study the doping dependence, as shown in Fig.~\ref{fig:stm_noproj}. We find that sFL evolves into the FL phase also upon increasing $x$: $B$ onsets at a small $x$ and increases with $x$, while $\Delta$ decreases with $x$.  For most of the doping range, they system remains on the sFL side, and both the pseudogap and the superconducting gap can be observed.  Interestingly, the superconductor gap shows a dome-like structure and reaches the maximal value at optimal doping $x_p\approx 0.5$.  In our theory, the pseudogap decreases monotonically with $x$.

\textbf{Nodal $p_x$ pairing} Below the superconducting gap, the tunneling spectrum shows a V-shape around $\omega=0$ and indicates a nodal pairing.  In Fig.~\ref{fig:stm_noproj}(c)(d), 
we show evidence of $p_x$ pairing around each of the two Fermi surfaces of the sFL phase.  Although the pairing of the spinon $\psi'$ is always on-site, the transferred pairing on the small hole pockets acquires a $p_x$ structure due to the winding of the spinor in the $(s_+, s_-)$ basis around the Fermi surface.

\section*{Discussion}

We propose the normal state of TBG above $T_c$ to be the sFL phase.  Here the slave boson condensation $B=0$, but the pairing gap of the spinon persists.  We assume that there is fluctuation of $B$ even above $T_c$, so the pairing gap from the spinon $\psi'$ loses coherence, but is still visible as a broadened peak at $\Delta_{\mathrm{PG}}\sim J\Delta$. Our theory thus offers a unified explanation of the pseudogap phase above $T_c$ and the two-gap SC phase below $T_c$ together.
A quantitative verification of this fluctuation-driven broadening lies beyond the present mean-field treatment and would require future Monte Carlo studies. 
There are a few remarks: (I) The major limitation of the current theory is that we ignored symmetry breaking orders such as the IKS state.  Therefore, the  theory is more suitable in the overdoped region, but may need modification at small $x$ to incorporate various IVC orders. (II) In the sFL phase, there should be a four-fold degeneracy in the Landau fan, in contrast to the two-fold observed in the experiment\cite{cao2018unconventional}. A small isospin polarization order is thus needed to match the experiments at finite field magnetic field. However, since magnetic field is not a perturbation for TBG, it is not clear whether such a symmetry breaking already exists at zero magnetic field. (III) We have assumed a spin-singlet pairing, but it is not necessarily in contradiction to the Pauli violation under in-plane field\cite{cao2021pauli}. The pairing of the spinons is at a much larger energy scale than $T_c$ and should be robust to Zeeman field.  The condensation of the slave boson $B$ is not directly from BCS pairing and may thus also persist to a Zeeman field well beyond the BCS Pauli  limit.

%\textbf{Conclusion} 
In summary, we propose a theory for superconductivity in TBG that emerges from an unconventional, symmetric normal state. This state, which we dub the ``second Fermi liquid" (sFL), features small hole pockets rather than the large Fermi surface of a conventional Fermi liquid. A key aspect of this sFL is the pre-existing pairing of local moments, mediated by optical phonons.  Upon lowering the temperature, the pre-formed pairing is transferred to the mobile carriers, opening a smaller superconducting gap on the small Fermi surfaces. Our theory provides a unified explanation for recent experimental observations of a  two-gap structure in the superconducting state and a pseudogap above $T_c$. 
We note that the pseudogap measurements were performed in magic-angle twisted trilayer graphene (TTG) \cite{park_pseudogap_2025,kim_twogap_2025}. 
Compared with TBG, TTG contains an additional Dirac cone at the $K$ point. 
However, the sFL-based superconductivity discussed here is associated with the $\Gamma$ point and is expected to remain essentially unchanged. 
Our theory also suggests that the TBG/TTG superconductivity emerges  close to the transition from sFL to the FL phase.  We propose to search for such a small to large Fermi surface transition in future experiments, for example, by tuning the twist angle.  We predict that the two gaps in the superconductor will merge to one single gap when entering the FL phase.

\textit{Note added:} When completing the manuscript, we became aware of a preprint \cite{wang_impurity_2025} that studied a single Kondo impurity model of the THFM and also identified a small Fermi surface state.

\section*{Methods}

The full THFM with interaction used in the main text is
\begin{equation}\label{eqn:THFM_0}
    H= H^{(c_1,c_2)}_0  + H_0^{(c_1, f)} + H_{\mathrm{int}}^{(f)}-\mu (N_c+N_f)~,
\end{equation}
which includes two dispersive bands, $c_1$ and $c_2$ described by $H_0^{(c_1,c_2)}$, and a flat  band $f$ described by the interaction $H_{\mathrm{int}}^{(f)}$. $f$ has a Wannier orbital well localized on the triangular lattice AA sites. 
The $c$ and $f$ orbitals are hybridized through $H_0^{(c_1,f)}$. 
%We use $f_{i}$ and $c_{1,\mathbf{k}}$, $c_{2,\mathbf{k}}$  as eight-component spinors, collecting spin, valley, and orbital flavor: $f_{i} = \{f_{i;\alpha}\}$, $c_{1,\mathbf{k}} = \{c_{1,\mathbf{k};\alpha}\}$ and $c_{2,\mathbf{k}} = \{c_{2,\mathbf{k};\alpha}\}$, with $\alpha=a \tau s$ formed by the orbital $a=\pm$, valley $\tau=K, K'$ and spin $s=\uparrow, \downarrow$. 
%Here we choose the orbital index $a=\pm$ such that $f_{a=\pm;\tau s}$ has angular momentum $L=\pm 1$ around each AA site.
Here 
\begin{equation}\label{eqn:THFM_c1c2}
\begin{aligned}
    H_0^{(c_1,c_2)} = &v_\star\sum_{\mathbf{k}}\left( c^\dagger_{1,\mathbf k}
    \tau_z\left( k_x\sigma_0+\mathrm{i}k_y\sigma_z\right) c_{2,\mathbf{k}}+\mathrm{h.c.}\right) \\
    &+\sum_{\mathbf{k}}c^\dagger_{2,\mathbf{k}}M\sigma_x c_{2,\mathbf{k}}, 
\end{aligned}
\end{equation}
\begin{equation}\label{eqn:THFM_c1f}
\begin{aligned}
    H^{(c_1,f)}_0=&\frac{1}{\sqrt{N}}\sum_{\mathbf{k},i} e^{\mathrm{i}\mathbf{k}\cdot\mathbf{R}_i-\frac{k^2\lambda^2}{2}}
    f^\dagger_{i}(\gamma\sigma_0 \\
    &+v^\prime_\star\tau_z\left(k_x\sigma_x+k_y\sigma_y\right)\big)
    c_{1,\mathbf{k}}+\mathrm{h.c.},
\end{aligned}
\end{equation}
\begin{equation}\label{eqn:interactionf}
\begin{aligned}
    H_{\mathrm{int}}^{(f)} %=& H_{U}^{(f)}+H_J^{(f)}  \\
    =U/2\sum_i(n_{i;f} - 4-\kappa\nu)^2 + \sum_ih_{i;J}^{(f)}, 
\end{aligned}
\end{equation}
where we use $\sigma_z,\tau_z,s_z$ to denote the Pauli matrices acting on the orbital, valley, and spin spaces, respectively. 
The band width of the itinerant $c_1,c_2$ band $v_\star |\mathbf{K}|$ is always the largest energy scale in the model, 
$M$ characterizes the $\Gamma$-point splitting of the active flat bands in TBG,  
and $\gamma$ characterizes the remote band gap at the $\Gamma$ point. 
%and $v'_\star$ characterize the hybridization between $c$ and $f$. 
We choose a set of parameters from \cite{Calugaru2023THFM,wang2024molecular} for a twist angle $\theta=1.06^\circ$ near the optimal doping, 
where we set $\gamma=-26.184$ meV, $v_\star = -4.335$ eV \AA,  $v'_\star = 1.633$ eV \AA\  and $\lambda = 0.339$ (in units of the moiré lattice constant). 
We follow Ref.~\cite{Lau2023THFM} to add a phenomenological parameter $-\kappa \nu$ in the Hubbard interaction,  
which approximates the remaining repulsion interaction between $c$ and $f$ orbitals at Hartree level. 
We choose $\kappa= 0.8$ throughout the paper. 
The value of the active band width $M$ mainly affects the nodal structure of the pairing state and is discussed in detail in Supplementary material. 
We take $M=0$ in the main text calculations. 

We follow Ref~\cite{wang2024molecular} to include two different kinds of effective interactions between the $f$ orbitals within each AA site. 
\begin{equation}\label{eqn:spinAH}
    h^{(f)}_{i;J} =  h^{(f)}_{i;J_A} +  h^{(f)}_{i;J_H}, 
\end{equation}
where $h^{(f)}_{i;J_A}$ is anti-Hund's coupling induced by the electron-phonon interaction, and  
$h^{(f)}_{i;J_H}$ is a Hund's coupling between different sublattices induced by the Coulomb interaction.
The explicit forms of $h^{(f)}_{i;J_A}$ and $h^{(f)}_{i;J_H}$ are provided and exactly solved in the supplementary. 
The doubly occupied ground state can be either an $L_z=0$ paired state or a pair of degenerate $L_z=\pm 2$ states, depending on parameters. 
In the main text, we focus on the case where a small heterostrain selects the doubly occupied ground state to be 
$|\Delta_{i;d}\rangle =  \frac{1}{2}(f_{i;+K\uparrow}^\dagger f_{i;+K'\downarrow}^\dagger-
f_{i;+K\downarrow}^\dagger f_{i;+K'\uparrow}^\dagger
+f_{i;-K\uparrow}^\dagger f_{i;-K'\downarrow}^\dagger-
f_{i;-K\downarrow}^\dagger f_{i;-K'\uparrow}^\dagger)|0\rangle$,  
which satisfy the $C_{2z}T$ symmetry, but breaks the $C_3$ symmetry.

\textbf{Mean field theory} We can substitute the parton construction in Eq.~\eqref{eqn:frac} to the Hamiltonian and then perform mean field decoupling.  
Together, we obtain the following mean-field Hamiltonian: 
\begin{equation}\label{eqn:THFM_mf}
\begin{aligned}
    H_{\mathrm{MF}} =& H_0^{(c_1,c_2)} + H_{\mathrm{MF}}^\Delta + H_{\mathrm{MF}}^B + \sum_i(E_s-\lambda+\mu) n_{i;s} \\
    &+ \sum_i(E_t-\lambda-\mu)n_{i;t} - \mu N_c\\
    H_{\mathrm{MF}}^{\Delta} =& \sum_{\mathbf k,\mathbf G} c^\dagger_{1,\mathbf{k+G}}\gamma(\mathbf{k+G})(\Delta s^\dagger_{-\mathbf k}+\sqrt{3}\Delta^* t_{\mathbf k})+\mathrm{h.c.}\\
    &- 2J\Delta^* \psi'_{-\mathbf k}\psi'_{\mathbf k} + \mathrm{h.c.}\\
    H_{\mathrm{MF}}^B = & \sum_{\mathbf k,\mathbf G} c^\dagger_{1,\mathbf{k+G}}\gamma(\mathbf{k+G})(B_s \psi'_{\mathbf k} + \Delta_t \psi'^\dagger_{-\mathbf k})+\mathrm{h.c.}\\
    & \sum_{\mathbf k} B'_s s^\dagger_{\mathbf{k}}\psi'_{\mathbf k} + \Delta'_tt^\dagger_\mathbf{k} \psi'_{\mathbf k} 
    +\mathrm{h.c.} ~.
\end{aligned}
\end{equation}
Here $s_{\mathbf{-k}}$, $t_{\mathbf{k}}$ are eight-component spinors, $s_{-\mathbf{k}} = \{\alpha s_{-\mathbf{k};\bar\alpha}\}$, $t^\dagger_{\mathbf{k}} = \{t^\dagger_{\mathbf{k};\alpha}\}$, where 
 $t_{\mathbf{k};\alpha} = \frac{1}{2\sqrt{3}} \sum_{\beta} \beta t_{\mathbf{k};\alpha\beta\bar\beta}$. 
And similar for $\psi'_{\mathbf k} = \{\psi_{\mathbf{k};\alpha}'\}$ and $\psi'_{-\mathbf k}=\{\alpha\psi'_{-\bar \alpha}\}$.
Here the order parameters $B_s \propto \langle s_{\beta}^\dagger\psi'_{\beta}\rangle$, $\Delta_t \propto \langle t_{\beta}^\dagger\psi'_{\beta}\rangle$, 
$B'_s \propto \langle c^\dagger_\alpha \gamma_{\alpha\beta} \psi'_\beta\rangle$, $\Delta'_t \propto \beta \langle c^\dagger_\alpha \gamma_{\alpha\beta} \psi'^
\dagger_{\bar\beta}\rangle$ and $\Delta \propto \alpha \langle \psi'_{\bar\alpha}\psi'_{\alpha}\rangle$
are determined self-consistently. 
For convergence, the slave bosons $B_s,B'_s$ and $\Delta_t,\Delta'_t$ are multiplied by an empirical coefficient $\alpha_B=1/2$ in mean-field iteration, as detailed in the Supplementary Material. 
An additional Lagrange multiplier $\lambda$ is introduced as $-\lambda(n_{i;s}+n_{i;t}+n_{i;\psi'}/2-1)$ to satisfy the local constraint on average. A chemical potential $\mu$ is introduced to satisfy the electron density constraint: $n_c+n_t-n_s=-x$.

\textbf{Data Availability} 
The code used to generate the results presented in this study will be made publicly available through Code Ocean upon publication.

\bibliography{refs}

@article{Cao2018mott,
    archivePrefix = {arXiv},
    arxivId = {1802.00553},
    author = {Cao, Yuan and Fatemi, Valla and Demir, Ahmet and Fang, Shiang and Tomarken, Spencer L. and Luo, Jason Y. and Sanchez-Yamagishi, Javier D. and Watanabe, Kenji and Taniguchi, Takashi and Kaxiras, Efthimios and Ashoori, Ray C. and Jarillo-Herrero, Pablo},
    doi = {10.1038/nature26154},
    eprint = {1802.00553},
    issn = {0028-0836},
    journal = {Nature},
    month = {apr},
    number = {7699},
    pages = {80--84},
    pmid = {29512654},
    publisher = {Nature Publishing Group},
    title = {{Correlated insulator behaviour at half-filling in magic-angle graphene superlattices}},
    volume = {556},
    year = {2018}
}

@article{wu2019phonon,
  title={Phonon-induced giant linear-in-T resistivity in magic angle twisted bilayer graphene: Ordinary strangeness and exotic superconductivity},
  doi={10.1103/PhysRevB.99.165112},
  author={Wu, Fengcheng and Hwang, Euyheon and Das Sarma, Sankar},
  journal={Physical Review B},
  volume={99},
  number={16},
  pages={165112},
  year={2019},
  publisher={APS}
}

@article{oh2024high,
  title={High-temperature superconductivity from kinetic energy},
  author={Oh, Hanbit and Yang, Hui and Zhang, Ya-Hui},
  journal={arXiv preprint arXiv:2411.07292},
  year={2024}
}

@article{senthil2000z,
  title={ ${Z_2}$ gauge theory of electron fractionalization in strongly correlated systems},
  doi={10.1103/PhysRevB.62.7850},
  author={Senthil, T and Fisher, Matthew PA},
  journal={Physical Review B},
  volume={62},
  number={12},
  pages={7850},
  year={2000},
  publisher={APS}
}

@article{senthil2003fractionalized,
  title={Fractionalized fermi liquids},
  doi={10.1103/PhysRevLett.90.216403},
  author={Senthil, T and Sachdev, Subir and Vojta, Matthias},
  journal={Physical review letters},
  volume={90},
  number={21},
  pages={216403},
  year={2003},
  publisher={APS}
}

@article{anderson1987resonating,
  title={The resonating valence bond state in La2CuO4 and superconductivity},
  author={Anderson, Philip W},
  journal={science},
  volume={235},
  number={4793},
  pages={1196--1198},
  year={1987},
  publisher={American Association for the Advancement of Science}
}

@article{kwan2021kekule,
  title={Kekul{\'e} spiral order at all nonzero integer fillings in twisted bilayer graphene},
  author={Kwan, Yves H and Wagner, Glenn and Soejima, Tomohiro and Zaletel, Michael P and Simon, Steven H and Parameswaran, Siddharth A and Bultinck, Nick},
  journal={Physical Review X},
  volume={11},
  number={4},
  doi={10.1103/PhysRevX.11.041063},
  pages={041063},
  year={2021},
  publisher={APS}
}

@article{khalaf2021charged,
  title={Charged skyrmions and topological origin of superconductivity in magic-angle graphene},
  doi={10.1126/sciadv.abf5299},
  author={Khalaf, Eslam and Chatterjee, Shubhayu and Bultinck, Nick and Zaletel, Michael P and Vishwanath, Ashvin},
  journal={Science advances},
  volume={7},
  number={19},
  pages={eabf5299},
  year={2021},
  publisher={American Association for the Advancement of Science}
}

@article{you2019superconductivity,
  title={Superconductivity from valley fluctuations and approximate $\mathrm{SO(4)}$ symmetry in a weak coupling theory of twisted bilayer graphene},
  doi={10.1038/s41535-019-0153-4},
  author={You, Yi-Zhuang and Vishwanath, Ashvin},
  journal={npj Quantum Materials},
  volume={4},
  number={1},
  pages={16},
  year={2019},
  publisher={Nature Publishing Group UK London}
}

@article{chou2019superconductor,
  title={Superconductor versus insulator in twisted bilayer graphene},
  doi={10.1103/PhysRevB.100.115128},
  author={Chou, Yang-Zhi and Lin, Yu-Ping and Das Sarma, Sankar and Nandkishore, Rahul M},
  journal={Physical Review B},
  volume={100},
  number={11},
  pages={115128},
  year={2019},
  publisher={APS}
}

@article{lian2019twisted,
  title={Twisted bilayer graphene: a phonon-driven superconductor},
  doi={10.1103/PhysRevLett.122.257002},
  author={Lian, Biao and Wang, Zhijun and Bernevig, B Andrei},
  journal={Physical review letters},
  volume={122},
  number={25},
  pages={257002},
  year={2019},
  publisher={APS}
}

@article{Yankowitz2019layer,
    author = {Yankowitz, Matthew and Chen, Shaowen and Polshyn, Hryhoriy and Zhang, Yuxuan and Watanabe, K. and Taniguchi, T. and Graf, David and Young, Andrea F. and Dean, Cory R.},
    doi = {10.1126/science.aav1910},
    issn = {0036-8075},
    journal = {Science},
    month = {mar},
    number = {6431},
    pages = {1059--1064},
    title = {{Tuning superconductivity in twisted bilayer graphene}},
    volume = {363},
    year = {2019}
}

@article{Lu2019moreSC,
    archivePrefix = {arXiv},
    arxivId = {1903.06513},
    author = {Lu, Xiaobo and Stepanov, Petr and Yang, Wei and Xie, Ming and Aamir, Mohammed Ali and Das, Ipsita and Urgell, Carles and Watanabe, Kenji and Taniguchi, Takashi and Zhang, Guangyu and Bachtold, Adrian and MacDonald, Allan H. and Efetov, Dmitri K.},
    doi = {10.1038/s41586-019-1695-0},
    eprint = {1903.06513},
    isbn = {4158601916},
    issn = {0028-0836},
    journal = {Nature},
    month = {oct},
    number = {7780},
    pages = {653--657},
    pmid = {31666722},
    publisher = {Springer US},
    title = {{Superconductors, orbital magnets and correlated states in magic-angle bilayer graphene}},
    volume = {574},
    year = {2019}
}

@article{Stepanov2020,
    archivePrefix = {arXiv},
    arxivId = {1911.09198},
    author = {Stepanov, Petr and Das, Ipsita and Lu, Xiaobo and Fahimniya, Ali and Watanabe, Kenji and Taniguchi, Takashi and Koppens, Frank H. L. and Lischner, Johannes and Levitov, Leonid and Efetov, Dmitri K.},
    doi = {10.1038/s41586-020-2459-6},
    eprint = {1911.09198},
    issn = {0028-0836},
    journal = {Nature},
    month = {jul},
    number = {7816},
    pages = {375--378},
    pmid = {32632215},
    publisher = {Springer US},
    title = {{Untying the insulating and superconducting orders in magic-angle graphene}},
    volume = {583},
    year = {2020}
}

@article{Cao2021nematic,
    archivePrefix = {arXiv},
    arxivId = {2004.04148},
    author = {Cao, Yuan and Rodan-Legrain, Daniel and Park, Jeong Min and Yuan, Noah F. Q. and Watanabe, Kenji and Taniguchi, Takashi and Fernandes, Rafael M. and Fu, Liang and Jarillo-Herrero, Pablo},
    doi = {10.1126/science.abc2836},
    eprint = {2004.04148},
    issn = {0036-8075},
    journal = {Science},
    month = {apr},
    number = {6539},
    pages = {264--271},
    pmid = {33859029},
    title = {{Nematicity and competing orders in superconducting magic-angle graphene}},
    volume = {372},
    year = {2021}
}

@article{Liu2021Coulomb,
    archivePrefix = {arXiv},
    arxivId = {2003.11072},
    author = {Liu, Xiaoxue and Wang, Zhi and Watanabe, K and Taniguchi, T and Vafek, Oskar and Li, J. I. A.},
    doi = {10.1126/science.abb8754},
    eprint = {2003.11072},
    issn = {0036-8075},
    journal = {Science},
    month = {mar},
    number = {6535},
    pages = {1261--1265},
    title = {{Tuning electron correlation in magic-angle twisted bilayer graphene using Coulomb screening}},
    volume = {371},
    year = {2021}
}

@article{huang_pseudospin_2022,
	title = {Pseudospin {Paramagnons} and the {Superconducting} {Dome} in {Magic} {Angle} {Twisted} {Bilayer} {Graphene}},
	volume = {129},
	issn = {0031-9007, 1079-7114},
	doi = {10.1103/PhysRevLett.129.187001},
	number = {18},
	urldate = {2025-10-31},
	journal = {Physical Review Letters},
	author = {Huang, Chunli and Wei, Nemin and Qin, Wei and MacDonald, Allan},
	month = oct,
	year = {2022},
	pages = {187001}
}

@article{wang2024molecular,
  title={Molecular pairing in twisted bilayer graphene superconductivity},
  author={Wang, Yi-Jie and Zhou, Geng-Dong and Peng, Shi-Yu and Lian, Biao and Song, Zhi-Da},
  journal={Physical Review Letters},
  volume={133},
  number={14},
  pages={146001},
  doi={10.1103/PhysRevLett.133.146001},
  year={2024},
  publisher={APS}
}

@article{chen2024strong,
  title={Strong electron--phonon coupling in magic-angle twisted bilayer graphene},
  author={Chen, Cheng and Nuckolls, Kevin P and Ding, Shuhan and Miao, Wangqian and Wong, Dillon and Oh, Myungchul and Lee, Ryan L and He, Shanmei and Peng, Cheng and Pei, Ding and others},
  journal={Nature},
  volume={636},
  number={8042},
  doi={10.1038/s41586-024-08227-w},
  pages={342--347},
  year={2024},
  publisher={Nature Publishing Group}
}

@misc{park_pseudogap_2025,
    archivePrefix = {arXiv},
    arxivId = {2503.16410},
    eprint = {2503.16410},
	title = {Simultaneous transport and tunneling spectroscopy of moiré graphene: {Distinct} observation of the superconducting gap and signatures of nodal superconductivity},
	doi = {10.48550/arXiv.2503.16410},
	publisher = {arXiv},
	author = {Park, Jeong Min and Sun, Shuwen and Watanabe, Kenji and Taniguchi, Takashi and Jarillo-Herrero, Pablo},
	month = mar,
	year = {2025}
}

@misc{kim_twogap_2025,
    archivePrefix = {arXiv},
    arxivId = {2505.17200},
    eprint = {2505.17200},
	title = {Resolving {Intervalley} {Gaps} and {Many}-{Body} {Resonances} in {Moiré} {Superconductor}},
	doi = {10.48550/arXiv.2505.17200},
	author = {Kim, Hyunjin and Rai, Gautam and Crippa, Lorenzo and Călugăru, Dumitru and Hu, Haoyu and Choi, Youngjoon and Kong, Lingyuan and Baum, Eli and Zhang, Yiran and Holleis, Ludwig and Watanabe, Kenji and Taniguchi, Takashi and Young, Andrea F. and Bernevig, B. Andrei and Valentí, Roser and Sangiovanni, Giorgio and Wehling, Tim and Nadj-Perge, Stevan},
	month = may,
	year = {2025}
}

@article{Bistritzer2011,
    archivePrefix = {arXiv},
    arxivId = {1009.4203},
    author = {Bistritzer, Rafi and MacDonald, Allan H.},
    doi = {10.1073/pnas.1108174108},
    eprint = {1009.4203},
    isbn = {1108174108},
    issn = {0027-8424},
    journal = {Proc. Natl. Acad. Sci.},
    month = {jul},
    number = {30},
    pages = {12233--12237},
    pmid = {21730173},
    title = {{Moir{\'{e}} bands in twisted double-layer graphene}},
    volume = {108},
    year = {2011}
}

@article{Po2018IVC,
    archivePrefix = {arXiv},
    arxivId = {1803.09742},
    author = {Po, Hoi Chun and Zou, Liujun and Vishwanath, Ashvin and Senthil, T},
    doi = {10.1103/PhysRevX.8.031089},
    eprint = {1803.09742},
    issn = {2160-3308},
    journal = {Physical Review X},
    month = {sep},
    number = {3},
    pages = {031089},
    title = {{Origin of Mott Insulating Behavior and Superconductivity in Twisted Bilayer Graphene}},
    volume = {8},
    year = {2018}
}

@article{Tarnopolsky2019,
    archivePrefix = {arXiv},
    arxivId = {2111.10018},
    author = {Tarnopolsky, Grigory and Kruchkov, Alex Jura and Vishwanath, Ashvin},
    doi = {10.1103/PhysRevLett.122.106405},
    eprint = {2111.10018},
    issn = {0031-9007},
    journal = {Physical Review Letter},
    month = {mar},
    number = {10},
    pages = {106405},
    publisher = {American Physical Society},
    title = {{Origin of Magic Angles in Twisted Bilayer Graphene}},
    volume = {122},
    year = {2019}
}

@article{Ahn2019fgtop,
    author = {Ahn, Junyeong and Park, Sungjoon and Yang, Bohm-jung},
    doi = {10.1103/PhysRevX.9.021013},
    issn = {2160-3308},
    journal = {Physical Review X},
    month = {apr},
    number = {2},
    pages = {021013},
    publisher = {American Physical Society},
    title = {{Failure of Nielsen-Ninomiya Theorem and Fragile Topology in Two-Dimensional Systems with Space-Time Inversion Symmetry: Application to Twisted Bilayer Graphene at Magic Angle}},
    volume = {9},
    year = {2019}
}

@article{Song2021symano,
    author = {Song, Zhi-da and Lian, Biao and Regnault, Nicolas and Bernevig, B Andrei},
    doi = {10.1103/PhysRevB.103.205412},
    issn = {2469-9950},
    journal = {Physical Review B},
    month = {may},
    number = {20},
    pages = {205412},
    publisher = {American Physical Society},
    title = {{Twisted bilayer graphene. II. Stable symmetry anomaly}},
    volume = {103},
    year = {2021}
}

@article{Ledwith2021Peda,
    archivePrefix = {arXiv},
    arxivId = {2105.08858},
    author = {Ledwith, Patrick J. and Khalaf, Eslam and Vishwanath, Ashvin},
    doi = {10.1016/j.aop.2021.168646},
    eprint = {2105.08858},
    issn = {00034916},
    journal = {Annals of Physics},
    month = {dec},
    pages = {168646},
    title = {{Strong coupling theory of magic-angle graphene: A pedagogical introduction}},
    volume = {435},
    year = {2021}
}

@article{Ledwith2024,
    archivePrefix = {arXiv},
    arxivId = {2408.16761},
    author = {Ledwith, Patrick J. and Dong, Junkai and Vishwanath, Ashvin and Khalaf, Eslam},
    eprint = {2408.16761},
    journal={arXiv preprint},
    month = {aug},
    pages = {1},
    title = {{Nonlocal Moments in the Chern Bands of Twisted Bilayer Graphene}},
    year = {2024}
}

@article{Song2022THFM,
    archivePrefix = {arXiv},
    arxivId = {2111.05865},
    author = {Song, Zhi-Da and Bernevig, B. Andrei},
    doi = {10.1103/PhysRevLett.129.047601},
    eprint = {2111.05865},
    isbn = {2100006118},
    issn = {0031-9007},
    journal = {Physical Review Letter},
    month = {jul},
    number = {4},
    pages = {047601},
    title = {{Magic-Angle Twisted Bilayer Graphene as a Topological Heavy Fermion Problem}},
    volume = {129},
    year = {2022}
}

@article{Calugaru2023THFM,
    archivePrefix = {arXiv},
    arxivId = {2303.03429},
    author = {Călugăru, Dumitru and Borovkov, Maksim and Lau, Liam L H and Coleman, Piers and Song, Zhi-da and Bernevig, B Andrei},
    doi = {10.1063/10.0019421},
    eprint = {2303.03429},
    issn = {1063-777X},
    journal = {Low Temperature Physics},
    month = {jun},
    number = {6},
    pages = {640--654},
    title = {{Twisted bilayer graphene as topological heavy fermion: II. Analytical approximations of the model parameters}},
    volume = {49},
    year = {2023}
}

@article{Yu2023THFM,
    author = {Yu, Jiabin and Xie, Ming and Bernevig, B Andrei and {Das Sarma}, Sankar},
    doi = {10.1103/PhysRevB.108.035129},
    issn = {2469-9950},
    journal = {Physical Review B},
    month = {jul},
    number = {3},
    pages = {035129},
    publisher = {American Physical Society},
    title = {{Magic-angle twisted symmetric trilayer graphene as a topological heavy-fermion problem}},
    volume = {108},
    year = {2023}
}

@article{Hu2023THFM,
    author = {Hu, Haoyu and Bernevig, B. Andrei and Tsvelik, Alexei M.},
    doi = {10.1103/PhysRevLett.131.026502},
    issn = {0031-9007},
    journal = {Physical Review Letters},
    month = {jul},
    number = {2},
    pages = {026502},
    publisher = {American Physical Society},
    title = {{Kondo Lattice Model of Magic-Angle Twisted-Bilayer Graphene: Hund's Rule, Local-Moment Fluctuations, and Low-Energy Effective Theory}},
    volume = {131},
    year = {2023}
}

@article{Vafek2024THFM,
    archivePrefix = {arXiv},
    arxivId = {2404.07253},
    author = {Herzog-Arbeitman, Jonah and Yu, Jiabin and Călugăru, Dumitru and Hu, Haoyu and Regnault, Nicolas and Liu, Chaoxing and Vafek, Oskar and Coleman, Piers and Tsvelik, Alexei and Song, Zhi-da and Bernevig, B. Andrei},
    eprint = {2404.07253},
    month = {apr},
    pages = {1--58},
    journal = {arXiv preprint},
    title = {{Topological Heavy Fermion Principle For Flat (Narrow) Bands With Concentrated Quantum Geometry}},
    url = {http://arxiv.org/abs/2404.07253},
    year = {2024}
}

@article{Zhou2024THFM,
    author = {Zhou, Geng-Dong and Wang, Yi-Jie and Tong, Ninghua and Song, Zhi-Da},
    doi = {10.1103/PhysRevB.109.045419},
    issn = {2469-9950},
    journal = {Physical Review B},
    month = {jan},
    number = {4},
    pages = {045419},
    publisher = {American Physical Society},
    title = {{Kondo phase in twisted bilayer graphene}},
    volume = {109},
    year = {2024}
}

@article{Hu2023THFM2,
    archivePrefix = {arXiv},
    arxivId = {2301.04673},
    author = {Hu, Haoyu and Rai, Gautam and Crippa, Lorenzo and Herzog-Arbeitman, Jonah and Călugăru, Dumitru and Wehling, Tim and Sangiovanni, Giorgio and Valent{\'{i}}, Roser and Tsvelik, Alexei M. and Bernevig, B. Andrei},
    doi = {10.1103/PhysRevLett.131.166501},
    eprint = {2301.04673},
    issn = {0031-9007},
    journal = {Physical Review Letters},
    month = {oct},
    number = {16},
    pages = {166501},
    pmid = {37925696},
    title = {{Symmetric Kondo Lattice States in Doped Strained Twisted Bilayer Graphene}},
    volume = {131},
    year = {2023}
}

@article{Chou2023THFM,
    archivePrefix = {arXiv},
    arxivId = {2211.15682},
    author = {Chou, Yang-Zhi and {Das Sarma}, Sankar},
    doi = {10.1103/PhysRevLett.131.026501},
    eprint = {2211.15682},
    issn = {0031-9007},
    journal = {Physical Review Letters},
    month = {jul},
    number = {2},
    pages = {026501},
    pmid = {37505969},
    publisher = {American Physical Society},
    title = {{Kondo Lattice Model in Magic-Angle Twisted Bilayer Graphene}},
    volume = {131},
    year = {2023}
}

@article{Wang2024THFM,
    archivePrefix = {arXiv},
    arxivId = {2407.11116},
    author = {Wang, Yi-Jie and Zhou, Geng-Dong and Lian, Biao and Song, Zhi-Da},
    eprint = {2407.11116},
    journal = {arXiv preprint},
    month = {jul},
    pages = {1--54},
    title = {{Electron phonon coupling in the topological heavy fermion model of twisted bilayer graphene}},
    year = {2024}
}

@article{Lau2023THFM,
    archivePrefix = {arXiv},
    arxivId = {2303.02670},
    author = {Lau, Liam L. H. and Coleman, Piers},
    eprint = {2303.02670},
    journal = {arXiv preprint},
    month = {mar},
    pages = {1--28},
    title = {{Topological Mixed Valence Model for Twisted Bilayer Graphene}},
    year = {2023}
}

@article{Rai2024THFM,
    author = {Rai, Gautam and Crippa, Lorenzo and Călugăru, Dumitru and Hu, Haoyu and Paoletti, Francesca and de' Medici, Luca and Georges, Antoine and Bernevig, B. Andrei and Valent{\'{i}}, Roser and Sangiovanni, Giorgio and Wehling, Tim},
    doi = {10.1103/PhysRevX.14.031045},
    issn = {2160-3308},
    journal = {Physical Review X},
    month = {sep},
    number = {3},
    pages = {031045},
    title = {{Dynamical Correlations and Order in Magic-Angle Twisted Bilayer Graphene}},
    volume = {14},
    year = {2024}
}

@article{Youn2024DMFT,
    archivePrefix = {arXiv},
    arxivId = {2412.03108},
    author = {Youn, Seongyeon and Goh, Beomjoon and Zhou, Geng-Dong and Song, Zhi-Da and Lee, Seung-Sup B.},
    eprint = {2412.03108},
    journal = {arXiv preprint},
    number = {mod 3},
    pages = {1--10},
    title = {{Hundness in twisted bilayer graphene: correlated gaps and pairing}},
    volume = {1},
    year = {2024}
}

@article{Hu2025THFM,
    archivePrefix = {arXiv},
    arxivId = {2502.14039},
    author = {Hu, Haoyu and Song, Zhi-Da and Bernevig, B. Andrei},
    eprint = {2502.14039},
    month = {feb},
    pages = {225--227},
    journal = {arXiv preprint},
    title = {{Projected and Solvable Topological Heavy Fermion Model of Twisted Bilayer Graphene}},
    year = {2025}
}

@misc{wang_impurity_2025,
    archivePrefix = {arXiv},
    arxivId = {2510.23604},
    eprint = {2510.23604},
	title = {Solution to a {Quantum} {Impurity} {Model} for {Moiré} {Systems}: {Fermi} {Liquid}, {Pairing}, and {Pseudogap}},
	author = {Wang, Yi-Jie and Zhou, Geng-Dong and Jung, Hyunsung and Youn, Seongyeon and Lee, Seung-Sup B. and Song, Zhi-Da},
	month = oct,
	year = {2025}
}

@article{Zhang2020,
    archivePrefix = {arXiv},
    arxivId = {2001.09159},
    author = {Zhang, Ya-Hui and Sachdev, Subir},
    doi = {10.1103/PhysRevResearch.2.023172},
    eprint = {2001.09159},
    issn = {2643-1564},
    journal = {Physical Review Research},
    month = {may},
    number = {2},
    pages = {023172},
    publisher = {American Physical Society},
    title = {{From the pseudogap metal to the Fermi liquid using ancilla qubits}},
    volume = {2},
    year = {2020}
}

@article{zhao_tbg_2025,
	title = {Topological {Mott} localization and pseudogap metal in twisted bilayer graphene},
	volume = {112},
	issn = {2469-9950, 2469-9969},
	doi = {10.1103/9n8v-7rx2},
	number = {8},
	urldate = {2025-10-06},
	journal = {Physical Review B},
	author = {Zhao, Jing-Yu and Zhou, Boran and Zhang, Ya-Hui},
	month = aug,
	year = {2025},
	pages = {085107}
}

@article{zhao_mixed_2025,
	title = {Mixed valence {Mott} insulator and composite excitation in twisted bilayer graphene},
	doi = {10.48550/arXiv.2507.00139},
    arXivId = {2507.00139},
    archivePrefix = {arXiv},
    eprint = {2507.00139},
	publisher = {arXiv},
    journal={arXiv preprint},
	author = {Zhao, Jing-Yu and Zhou, Boran and Zhang, Ya-Hui},
	month = jun,
	year = {2025}
}

@article{andrei2020graphene,
  title={Graphene bilayers with a twist},
  author={Andrei, Eva Y and MacDonald, Allan H},
  journal={Nature materials},
  volume={19},
  number={12},
  pages={1265--1275},
  year={2020},
  doi={10.1038/s41563-020-00917-w},
  publisher={Nature Publishing Group UK London}
}

@article{andrei2021marvels,
  title={The marvels of moir{\'e} materials},
  author={Andrei, Eva Y and Efetov, Dmitri K and Jarillo-Herrero, Pablo and MacDonald, Allan H and Mak, Kin Fai and Senthil, T and Tutuc, Emanuel and Yazdani, Ali and Young, Andrea F},
  journal={Nature Reviews Materials},
  volume={6},
  doi={10.1038/s41578-021-00284-1},
  number={3},
  pages={201--206},
  year={2021},
  publisher={Nature Publishing Group UK London}
}

@article{nuckolls2024microscopic,
  title={A microscopic perspective on moir{\'e} materials},
  author={Nuckolls, Kevin P and Yazdani, Ali},
  journal={Nature Reviews Materials},
  volume={9},
  number={7},
  pages={460--480},
  year={2024},
  doi={10.1038/s41578-024-00682-1},
  publisher={Nature Publishing Group UK London}
}

@article{bultinck2020ground,
  title={Ground state and hidden symmetry of magic-angle graphene at even integer filling},
  author={Bultinck, Nick and Khalaf, Eslam and Liu, Shang and Chatterjee, Shubhayu and Vishwanath, Ashvin and Zaletel, Michael P},
  journal={Physical Review X},
  volume={10},
  number={3},
  doi={10.1103/PhysRevX.10.031034},
  pages={031034},
  year={2020},
  publisher={APS}
}

@article{parker2021strain,
  title={Strain-induced quantum phase transitions in magic-angle graphene},
  author={Parker, Daniel E and Soejima, Tomohiro and Hauschild, Johannes and Zaletel, Michael P and Bultinck, Nick},
  journal={Physical review letters},
  volume={127},
  doi={10.1103/PhysRevLett.127.027601},
  number={2},
  pages={027601},
  year={2021},
  publisher={APS}
}

@article{wagner2022global,
  title={Global phase diagram of the normal state of twisted bilayer graphene},
  author={Wagner, Glenn and Kwan, Yves H and Bultinck, Nick and Simon, Steven H and Parameswaran, SA},
  journal={Physical review letters},
  volume={128},
  number={15},
  pages={156401},
  doi={10.1103/PhysRevLett.128.156401},
  year={2022},
  publisher={APS}
}

@article{rozen2021entropic,
  title={Entropic evidence for a Pomeranchuk effect in magic-angle graphene},
  author={Rozen, Asaf and Park, Jeong Min and Zondiner, Uri and Cao, Yuan and Rodan-Legrain, Daniel and Taniguchi, Takashi and Watanabe, Kenji and Oreg, Yuval and Stern, Ady and Berg, Erez and others},
  journal={Nature},
  volume={592},
  number={7853},
  doi={10.1038/s41586-021-03319-3},
  pages={214--219},
  year={2021},
  publisher={Nature Publishing Group UK London}
}

@article{saito2021isospin,
  title={Isospin Pomeranchuk effect in twisted bilayer graphene},
  author={Saito, Yu and Yang, Fangyuan and Ge, Jingyuan and Liu, Xiaoxue and Taniguchi, Takashi and Watanabe, Kenji and Li, JIA and Berg, Erez and Young, Andrea F},
  journal={Nature},
  volume={592},
  doi={10.1038/s41586-021-03409-2},
  number={7853},
  pages={220--224},
  year={2021},
  publisher={Nature Publishing Group UK London}
}

@article{zhang2020spin,
  title={Spin liquids and pseudogap metals in the SU (4) Hubbard model in a moir{\'e} superlattice},
  author={Zhang, Ya-Hui and Mao, Dan},
  journal={Physical Review B},
  volume={101},
  number={3},
  pages={035122},
  year={2020},
  doi={10.1103/PhysRevB.101.035122},
  publisher={APS}
}

@article{yang2024strong,
  title={Strong pairing from a small Fermi surface beyond weak coupling: Application to $\mathrm{La}_3\mathrm{Ni}_2\mathrm{O}_7$},
  author={Yang, Hui and Oh, Hanbit and Zhang, Ya-Hui},
  journal={Physical Review B},
  volume={110},
  number={10},
  pages={104517},
  year={2024},
  doi={10.1103/PhysRevB.110.104517},
  publisher={APS}
}

@article{lee2006doping,
    archivePrefix = {arXiv},
    arxivId = {cond-mat/0410445},
    author = {Lee, Patrick A. and Nagaosa, Naoto and Wen, Xiao-Gang},
    doi = {10.1103/RevModPhys.78.17},
    eprint = {0410445},
    issn = {0034-6861},
    journal = {Reviews of Modern Physics},
    month = {jan},
    number = {1},
    pages = {17--85},
    title = {{Doping a Mott insulator: Physics of high-temperature superconductivity}},
    volume = {78},
    year = {2006}
}

@article{ledwith2025exotic,
  archivePrefix = {arXiv},
  arxivId = {2505.08779},
  eprint = {2505.08779},
  title={Exotic Carriers from Concentrated Topology: Dirac Trions as the Origin of the Missing Spectral Weight in Twisted Bilayer Graphene},
  author={Ledwith, Patrick J and Vishwanath, Ashvin and Khalaf, Eslam},
  journal={arXiv}, 
  year={2025}
}

@article{nuckolls2023quantum,
  title={Quantum textures of the many-body wavefunctions in magic-angle graphene},
  author={Nuckolls, Kevin P and Lee, Ryan L and Oh, Myungchul and Wong, Dillon and Soejima, Tomohiro and Hong, Jung Pyo and C{\u{a}}lug{\u{a}}ru, Dumitru and Herzog-Arbeitman, Jonah and Bernevig, B Andrei and Watanabe, Kenji and others},
  journal={Nature},
  volume={620},
  number={7974},
  pages={525--532},
  doi={10.1038/s41586-023-06226-x},
  year={2023},
  publisher={Nature Publishing Group UK London}
}

@article{cao2018unconventional,
  title={Unconventional superconductivity in magic-angle graphene superlattices},
  author={Cao, Yuan and Fatemi, Valla and Fang, Shiang and Watanabe, Kenji and Taniguchi, Takashi and Kaxiras, Efthimios and Jarillo-Herrero, Pablo},
  journal={Nature},
  volume={556},
  number={7699},
  pages={43--50},
  doi={doi.org/10.1038/nature26160},
  year={2018},
  publisher={Nature Publishing Group UK London}
}

@misc{wang_kekule_2025,
    archivePrefix = {arXiv},
    arxivId = {2510.06451},
    eprint = {2510.06451},
	title = {Kekulé {Superconductivity} in {Twisted} {Magic} {Angle} {Bilayer} {Graphene}},
	urldate = {2025-10-13},
	publisher = {arXiv},
	author = {Wang, Ke and Levin, K.},
	month = oct,
	year = {2025}
}

@misc{wang_IKS_2025,
    archivePrefix = {arXiv},
    arxivId = {2509.12320},
    eprint = {2509.12320},
	title = {Putting a new spin on the incommensurate {Kekulé} spiral: from spin-valley locking and collective modes to fermiology and implications for superconductivity},
	publisher = {arXiv},
	author = {Wang, Ziwei and Wagner, Glenn and Kwan, Yves H. and Bultinck, Nick and Simon, Steven H. and Parameswaran, S. A.},
	month = sep,
	year = {2025}
}

@article{wu_ephsc_2018,
	title = {Theory of {Phonon}-{Mediated} {Superconductivity} in {Twisted} {Bilayer} {Graphene}},
	volume = {121},
	doi = {10.1103/PhysRevLett.121.257001},
	number = {25},
	journal = {Physical Review Letters},
	author = {Wu, Fengcheng},
	year = {2018}
}

@article{sharma_plasma_2020,
	title = {Superconductivity from collective excitations in magic-angle twisted bilayer graphene},
	volume = {2},
	issn = {2643-1564},
	doi = {10.1103/PhysRevResearch.2.022040},
	number = {2},
	journal = {Physical Review Research},
	author = {Sharma, Gargee and Trushin, Maxim and Sushkov, Oleg P. and Vignale, Giovanni and Adam, Shaffique},
	month = may,
	year = {2020},
	pages = {022040}
}

@article{kennes_frg_2018,
	title = {Strong correlations and $d+id$ superconductivity in twisted bilayer graphene},
	volume = {98},
	doi = {10.1103/PhysRevB.98.241407},
	number = {24},
	journal = {Physical Review B},
	author = {Kennes, Dante M.},
	year = {2018}
}

@article{gonzalez_kohn_luttinger_2019,
	title = {Kohn-{Luttinger} {Superconductivity} in {Twisted} {Bilayer} {Graphene}},
	volume = {122},
	issn = {0031-9007, 1079-7114},
	url = {http://arxiv.org/abs/1807.01275},
	doi = {10.1103/PhysRevLett.122.026801},
	number = {2},
	urldate = {2025-10-18},
	journal = {Physical Review Letters},
	author = {González, J. and Stauber, T.},
	month = jan,
	year = {2019},
	note = {arXiv:1807.01275 [cond-mat]},
	pages = {026801}
}

@article{isobe_vanHove_2018,
	title = {Unconventional {Superconductivity} and {Density} {Waves} in {Twisted} {Bilayer} {Graphene}},
	volume = {8},
	issn = {2160-3308},
	doi = {10.1103/PhysRevX.8.041041},
	number = {4},
	urldate = {2025-10-18},
	journal = {Physical Review X},
	author = {Isobe, Hiroki and Yuan, Noah F. Q. and Fu, Liang},
	month = dec,
	year = {2018},
	pages = {041041}
}

@article{chichinadze_vanHove_2020,
	title = {Nematic superconductivity in twisted bilayer graphene},
	volume = {101},
	issn = {2469-9950, 2469-9969},
	doi = {10.1103/PhysRevB.101.224513},
	number = {22},
	urldate = {2025-10-18},
	journal = {Physical Review B},
	author = {Chichinadze, Dmitry V. and Classen, Laura and Chubukov, Andrey V.},
	month = jun,
	year = {2020},
	pages = {224513}
}

@article{kim2023imaging,
  title={Imaging inter-valley coherent order in magic-angle twisted trilayer graphene},
  author={Kim, Hyunjin and Choi, Youngjoon and Lantagne-Hurtubise, {\'E}tienne and Lewandowski, Cyprian and Thomson, Alex and Kong, Lingyuan and Zhou, Haoxin and Baum, Eli and Zhang, Yiran and Holleis, Ludwig and others},
  journal={Nature},
  volume={623},
  number={7989},
  doi={10.1038/s41586-023-06663-8},
  pages={942--948},
  year={2023},
  publisher={Nature Publishing Group UK London}
}

@article{kwan_kekule_2021,
	title = {Kekulé {Spiral} {Order} at {All} {Nonzero} {Integer} {Fillings} in {Twisted} {Bilayer} {Graphene}},
	volume = {11},
	issn = {2160-3308},
	doi = {10.1103/PhysRevX.11.041063},
	number = {4},
	urldate = {2025-10-01},
	journal = {Physical Review X},
	author = {Kwan, Y. H. and Wagner, G. and Soejima, T. and Zaletel, M. P. and Simon, S. H. and Parameswaran, S. A. and Bultinck, N.},
	month = dec,
	year = {2021},
	pages = {041063}
}

@article{park2021tunable,
  title={Tunable strongly coupled superconductivity in magic-angle twisted trilayer graphene},
  author={Park, Jeong Min and Cao, Yuan and Watanabe, Kenji and Taniguchi, Takashi and Jarillo-Herrero, Pablo},
  journal={Nature},
  volume={590},
  doi={10.1038/s41586-021-03192-0},
  number={7845},
  pages={249--255},
  year={2021},
  publisher={Nature Publishing Group UK London}
}

@article{hao2021electric,
  title={Electric field--tunable superconductivity in alternating-twist magic-angle trilayer graphene},
  author={Hao, Zeyu and Zimmerman, AM and Ledwith, Patrick and Khalaf, Eslam and Najafabadi, Danial Haie and Watanabe, Kenji and Taniguchi, Takashi and Vishwanath, Ashvin and Kim, Philip},
  journal={Science},
  volume={371},
  number={6534},
  doi={10.1126/science.abg0399},
  pages={1133--1138},
  year={2021},
  publisher={American Association for the Advancement of Science}
}

@article{park2022robust,
  title={Robust superconductivity in magic-angle multilayer graphene family},
  author={Park, Jeong Min and Cao, Yuan and Xia, Li-Qiao and Sun, Shuwen and Watanabe, Kenji and Taniguchi, Takashi and Jarillo-Herrero, Pablo},
  journal={Nature Materials},
  volume={21},
  number={8},
  pages={877--883},
  year={2022},
  doi={10.1038/s41563-022-01287-1},
  publisher={Nature Publishing Group UK London}
}

@article{arora2020superconductivity,
  title={Superconductivity in metallic twisted bilayer graphene stabilized by WSe2},
  author={Arora, Harpreet Singh and Polski, Robert and Zhang, Yiran and Thomson, Alex and Choi, Youngjoon and Kim, Hyunjin and Lin, Zhong and Wilson, Ilham Zaky and Xu, Xiaodong and Chu, Jiun-Haw and others},
  journal={Nature},
  volume={583},
  number={7816},
  pages={379--384},
  year={2020},
  doi={10.1038/s41586-020-2473-8},
  publisher={Nature Publishing Group UK London}
}

@article{cao2021pauli,
  title={Pauli-limit violation and re-entrant superconductivity in moir{\'e} graphene},
  author={Cao, Yuan and Park, Jeong Min and Watanabe, Kenji and Taniguchi, Takashi and Jarillo-Herrero, Pablo},
  journal={Nature},
  volume={595},
  number={7868},
  doi={10.1038/s41586-021-03685-y},
  pages={526--531},
  year={2021},
  publisher={Nature Publishing Group UK London}
}

\textbf{Acknowledgement}  
The work is supported by the Alfred P. Sloan Foundation through a Sloan Research Fellowship (Y.-H.Z.).

\textbf{Author contributions}
Y.-H. Z. conceived and supervised the project. 
J.-Y. Z. performed the theoretical and numerical computations described in the paper. 
All authors contributed to the interpretation of the results and writing of the manuscript.

\textbf{Competing interests}
The authors declare no competing interests.

\appendix

\begin{widetext}

\section{On-site spin interactions} \label{app:THFM}
%\subsection{Spin interaction}

Here we follow Ref~\cite{wang2024molecular} to consider two different kinds of effective interactions between the $f$ orbitals within each AA site. 
\begin{equation}\label{eqn:spinAH}
    h^{(f)}_{i;J} =  h^{(f)}_{i;J_A} +  h^{(f)}_{i;J_H}, 
\end{equation}
where 
\begin{equation}\label{eqn:antiHundH}
\begin{aligned}
    h_{i;J_A}^{(f)} =& -\frac{J_A}{2}\sum_{\alpha\eta\beta ss'} f^\dagger_{i;\beta\bar\eta s}f^\dagger_{i;\alpha\eta s'} f^{}_{i;\beta\bar\eta s'}f^{}_{i;\alpha\eta s}
    - \frac{J_A}{2}\sum_{\alpha\eta ss'} f^\dagger_{i;\alpha\bar\eta s}f^\dagger_{i;\bar\alpha \eta s'} f^{}_{i;\bar\alpha\bar\eta s'} f^{}_{i;\alpha\eta s} 
    %=&  \frac{J_A}{4}\sum_\eta N_\eta N_\eta +J_A\sum_\eta \mathbf{S}_\eta\cdot \mathbf{S}_\eta 
    %-\frac{J_A}{2}\sum_{\alpha\eta ss'} f^\dagger_{\alpha\bar\eta s}f^\dagger_{\bar\alpha \eta s'} f^{}_{\bar\alpha\bar\eta s'} f^{}_{\alpha\eta s}, 
\end{aligned}
\end{equation} 
is anti-Hund's coupling induced by the electron-phonon interaction. 
%where $\mu,\rho = x,y,z$ label Pauli matrices acting on orbital ($\sigma$) and spin ($s$) spaces, respectively.  
And 
%\begin{equation}\label{eqn:HundH}
%\begin{aligned}
    %h_{i;J_H}^{(f)} =
    %& \frac{J_H}{2} N + \sum_\alpha(\frac{J_H}{4} N^2_\alpha + \frac{J_H'}{4} N_\alpha N_\alpha - J_H S_\alpha^2 - J_H' S_\alpha\cdot S_\alpha) + 
    %\frac{J_H'}{2}\sum_{\alpha ss'\eta\eta'} f^\dagger_{\alpha\eta s}f^\dagger_{\alpha\bar\eta s'} f^{}_{\bar\alpha\bar\eta s'} f^{}_{\bar\alpha \eta' s}. 
%\end{aligned}
%\end{equation} 
\begin{equation}\label{eqn:HundH}
\begin{aligned}
h_{i;J_H}^{(f)} = & \sum_{\alpha s s'}  \sum_{\eta_{1,2,3,4}}
\delta_{\eta_1+\eta_2, \eta_3+\eta_4}
 \frac{J_H}{2} f^\dagger_{i;(\alpha\eta_1) \eta_1 s} f^\dagger_{i;(\alpha\eta_2) \eta_2 s'}
f_{i;(\alpha\eta_3) \eta_3 s'} f_{i;(\alpha\eta_4) \eta_4 s}\\
&+\sum_{\alpha ss'\eta_1,\eta_2} \Big[ \frac{J_H'}{2} 
f^\dagger_{i;(\alpha\eta_1) \eta_1 s} f^\dagger_{i;(\alpha\bar\eta_1) \bar\eta_1 s'}
f_{i;(\bar{\alpha}\bar\eta_2) \bar \eta_2 s'} f_{i;(\bar \alpha\eta_4) \eta_4 s}
+ \frac{J_H'}{2} f^\dagger_{i;(\alpha\eta_1) \eta_1 s} f^\dagger_{i;(\bar{\alpha}\eta_2) \eta_2 s'}
f_{i;\bar{\alpha} \eta_2 s'} f_{i;(\alpha\eta_1) \eta_1 s} 
\\&+ \frac{J_H'}{2} f^\dagger_{i;(\alpha\eta_1) \eta_1 s} f^\dagger_{i;(\bar{\alpha}\eta_2) \eta_2 s'}
f_{i;(\alpha\eta_1) \eta_1 s'} f_{i;(\bar{\alpha}\eta_2) \eta_2 s}
\Big].
\end{aligned}
\end{equation} 
is a Hund's coupling between different sublattices induced by the Coulomb interaction.

The spin interaction can be solved exactly for the singly, doubly and triply occupied states, respectively. 
The singly occupied states do not receive energy from the intra-site spin interaction, and the triply occupied state is not important in the doping regime $\nu=-2-x$ we studied here. 
Therefore, here we focus on the doubly occupied states, which  serves as the parent state. 
In combination of the above two terms, the on-site ground state is found to be an inter-valley singlet pairing state between either $a\eta s$ and $\bar a \bar\eta\bar s$, or between $a\eta s$ and $a\bar\eta\bar s$, depending on the detailed parameters. 
For $J_H'=J_H/3$ and $J_H<J_A/2$, the lowest energy state is found to be an $L_z=0$ state 
\begin{equation}
\begin{aligned}
    \ket{\Delta_{i;s}} = 
    &\frac{1}{2 }\left(f^\dagger_{i;+K\uparrow} f^\dagger_{i;-K'\downarrow} 
    - f^\dagger_{i;+K\downarrow} f^\dagger_{i;-K'\uparrow} \right.
    +\left.f^\dagger_{i;-K\uparrow} f^\dagger_{i;+K'\downarrow} 
    - f^\dagger_{i;-K\downarrow} f^\dagger_{i;+K'\uparrow} \right)\ket{0},\\
\end{aligned}
\end{equation}
In contrast, for $J_H'=J_H/3$ and $J_A<J_H <3J_A/2 $, the lowest-energy manifold is twofold degenerate, consisting of the states 
\begin{equation}
\begin{aligned}
    |\Delta_{i;d_1}\rangle = & \frac{1}{\sqrt{2}}(f_{i;+K\uparrow}^\dagger f_{i;+K'\downarrow}^\dagger-
    f_{i;+K\downarrow}^\dagger f_{i;+K'\uparrow}^\dagger)|0\rangle\\
    |\Delta_{i;d_2}\rangle = & \frac{1}{\sqrt{2}}(f_{i;-K\uparrow}^\dagger f_{i;-K'\downarrow}^\dagger-
    f_{i;-K\downarrow}^\dagger f_{i;-K'\uparrow}^\dagger)|0\rangle~.
\end{aligned}
\end{equation}
with angular momentum $L_z = \pm 2$, respectively. 
In our mean-field calculation, we will always assume an equal weight superposition of these two degenerate state
$|\Delta_{i;d}\rangle = \frac{1}{\sqrt{2}}(|\Delta_{i;d_1}\rangle + |\Delta_{i;d_2}\rangle)$, 
which satisfy the $C_{2z}T$ symmetry, but breaks the $C_3$ symmetry.  It should be selected  by a small finite heterostrain.

\section{Details on the mean field theory calculation}\label{app:MF}

\subsection{Slave particle theory}

To analytically treat the restricted Hilbert space which includes only singlon, doublon and triplon states of $f$ orbital, we propose the following parton construction:
\begin{equation}
\begin{aligned}
    f^\dagger_{i;\alpha}|0\rangle &\rightarrow s^\dagger_{i;\alpha}|0\rangle\\
    f^\dagger_{i;\alpha}f^\dagger_{i;\beta}|0\rangle &\rightarrow \psi'^\dagger_{i;\alpha}\psi'^\dagger_{i;\beta}|0\rangle\\
    f^\dagger_{i;\alpha}f^\dagger_{i;\beta}f^\dagger_{i;\gamma}|0\rangle &\rightarrow t^\dagger_{i;\alpha\beta\gamma}|0\rangle
\end{aligned}
\end{equation}
where $s$, $\psi'$ and $t$ are all fermions which satisfy a local constraint: 
\begin{equation}\label{eqn:constraint}
    n_{i;s}+n_{i;\psi'}/2+n_{i;t}=1, 
\end{equation}
at every AA site $i$. 
Here $n_{i;s}, n_{i;\psi'}$ and $n_{i;t}$ are the total particle numbers of $s,\psi'$ and $t$ at each site $i$. 
The constraint introduces a gauge redundancy $s_i\rightarrow s_ie^{2i\varphi_i}, \psi'_i\rightarrow \psi'_i e^{i\varphi_i}$ and $t_i\rightarrow t_ie^{2i\varphi_i}$.

We now write down the projected Hamiltonian $P_GH_0P_G$ in terms of these partons.
First, the original $f_i$ fermion operator is written as 
\begin{equation}\label{eqn:frac_app}
    P_Gf_{i;\alpha}^{}P_G = \sum_\beta s^\dagger_{i;\beta}\psi'_{i;\beta}\psi'_{i;\alpha}
    +\sum_{\beta,\gamma(\beta<\gamma)} t^{}_{i;\alpha\beta\gamma} \psi'^\dagger_{i;\beta}\psi'^\dagger_{i;\gamma}~.
\end{equation}
The bilinear part of the Hamiltonian $P_GH_0P_G$ can be obtained by directly replacing the $f^\dagger_{i;\alpha}$ operators with the expression above. 

The interaction energy \eqref{eqn:interactionf} are composed of two part.  
For the Hubbard interaction part, we write it down exactly in the restricted Hilbert space for each valences $f^{1+}, f^{2+}, f^{3+}$
as $E_{1+}^f = (3+\kappa\nu)^2U/2$, $E_{2+}^f=(2+\kappa\nu)^2U/2$ and $E_{3+}^f=(1+\kappa\nu)^2U/2$. 
The spin interaction is only meaningful for the doublon represented by $\psi'$. We thus replace the $f$ fermions in Eqs.~\eqref{eqn:antiHundH} and \eqref{eqn:HundH} with $\psi'$ as $h^{(f)}_{i;J}\rightarrow h_{i;J}^{(\psi')}$. 
We will keep the full interaction $h_{i;J}^{(\psi')}$ for now and leave it for the mean-field calculation later. 
With the chemical potential $\mu$ and a Lagrange multipler, the onsite interaction Hamiltonian can be written as:
\begin{equation}
\begin{aligned}
    &P_G (\sum_i H_{i;\mathrm{int}}^{(f)}-\mu N_f)P_G-\lambda_0 \sum_i(n_{i;}+n_{i;\psi'}/2+n_{i;t}-1) \\ 
    =&
    (E_s+\mu-\lambda) \sum_{i;\alpha} s^\dagger_{i;\alpha} s^{}_{i;\alpha} 
    -\frac{1}{2}\lambda\sum_{i;\alpha}\psi'^\dagger_{i;\alpha} \psi'_{i;\alpha} 
    + (E_t-\mu-\lambda)\sum_{i;\alpha\beta\gamma}t^\dagger _{i;\alpha\beta\gamma} t^{}_{i;\alpha\beta\gamma}~ + \sum_i h_{i;J}^{(\psi')} + \cdots~.%\mathrm{const.},
\end{aligned}
\end{equation}
where $\cdots$ denotes the omitted spin interaction within the triplon subspace, together with an overall constant.
We also shift the Lagrange multiplier to $\lambda_0 = E_{2+}^f - 2\mu + \lambda$ and 
\begin{equation}
    E_s = E_{1+}^f - E_{2+}^f = U/2 + U(2+\kappa\nu) , 
\end{equation}
\begin{equation}
    E_t = E_{3+}^f - E_{2+}^f = U/2 - U(2+\kappa\nu) . 
\end{equation}

In summary, the projected Hamiltonian written in terms of the partons is: 
\begin{equation}
\begin{aligned}
    P_GHP_G =& H_0^{(c_1,c_2)}  -\mu N_c  
    +\frac{1}{\sqrt{N}} \sum_{\mathbf k,\mathbf G,i}e^{i\mathbf k\cdot \mathbf R_i} c^\dagger_{1,\mathbf{k+G}}\gamma(\mathbf{k+G})\left(
    \sum_\beta s^\dagger_{i;\beta}\psi'_{i;\beta}\psi'_{i;\alpha}
    +\sum_{\beta\gamma(\beta<\gamma)} t^{}_{i;\alpha\beta\gamma} \psi'^\dagger_{i;\beta}\psi'^\dagger_{i;\gamma}
    \right)+\mathrm{h.c.}\\
    &+(E_s+\mu-\lambda) \sum_{i,\alpha} s^\dagger_{i;\alpha} s^{}_{i;\alpha} 
    -\frac{1}{2}\lambda\sum_{i,\alpha}\psi'^\dagger_{i;\alpha} \psi'_{i;\alpha} 
    + (E_t-\mu-\lambda)\sum_{i,\alpha}t^\dagger _{i;\alpha} t^{}_{i;\alpha}~ +\sum_{i} h_{i;J}^{(\psi')} + \mathrm{const.},
\end{aligned}
\end{equation}

\subsection{Mean field theory}

\subsubsection{Spin interaction}
In the mean-field calculation, we can decouple the full spin Hamiltonian Eq.~\eqref{eqn:spinAH} into both the $s$-wave and $d$-wave channels and solve the self-consistent equation. 
However, the $f_i$ orbitals are localized on the AA site and the intra-site spin Hamiltonian can be solved exactly.  
With knowledge of the ground state at different parameter regime, 
here we directly assume that the pairing channel is either the $s$-wave state $|\Delta_{i;s}\rangle$ or $C_{2z}T$ symmetric $d$-wave state $|\Delta_{i;d}\rangle$ and  consider the mean-field Hamiltonian
\begin{equation}
    H_{J,\mathrm{MF}}^s = -2J_s\sum_{i;a\eta s}\Delta^*_ss\psi'_{i;\bar a\bar\eta\bar s}\psi'_{i;a\eta s} + \mathrm{h.c.}~,
\end{equation}
for the $s$-wave pairing, where $\Delta_s = s\langle \psi'_{\bar a\bar\eta\bar s}\psi'_{a\eta s}\rangle$, $s=\pm$ is the spin singlet sign. 
And we consider 
\begin{equation}\label{eqn:mf_dwave}
    H_{J,\mathrm{MF}}^d = -2J_d\sum_{i;a\eta s}\Delta^*_ds\psi'_{i; a\bar\eta\bar s}\psi'_{i;a\eta s} + \mathrm{h.c.}~,
\end{equation}
for the $d$-wave pairing, with $\Delta_d = s\langle \psi'_{a\bar\eta\bar s}\psi'_{a\eta s}\rangle$. 
Here the value of $J_s$  or $J_d$ is generally a linear combination of the microscopic parameters $J_A$ and $J_H$.

Motivated by the experimental observation of a $V$-shaped spectrum, we choose the $d$-wave pairing channel as the pairing of the spinon $\psi'$ in the main text. Because we are not interested in the competition between the $d$ wave and $s$ wave ansatz, we can keep only one parameter $J=J_d$ for the spin interaction.
From now on, we use Eq.~\eqref{eqn:mf_dwave} and omit the subscript $d$ for simplicity.
Since each flavor $f_{i;\alpha}$ is paired uniquely with another flavor $f_{i;\bar\alpha}$ in $|\Delta_{d;i}\rangle$, we define
\begin{equation}
    \bar\alpha = a\bar\tau \bar s\quad \mathrm{for}\quad \alpha = a\tau s. 
\end{equation}
%And for $\alpha = a \eta s$, when $\alpha$ is not used as a subscript, we let $\alpha$ itself represent a sign $\pm$, determined by the spin $s = \uparrow,\downarrow$. 
By a slight abuse of notation, when $\alpha=a\eta s$ appears as a sign rather than as an index, we take $\alpha=+1$ for $s=\uparrow$ and $\alpha=-1$ for $s=\downarrow$.

\subsubsection{Hybridization term}

For the hybridization term $c^\dagger_{\mathbf{k};\alpha}\gamma(\mathbf{k})_{\alpha\beta} f_{i;\beta}$, we consider two different kinds of mean field channels. 
First when $J$ term is large, the $\psi'_{i;\alpha}$ particle would fall into a pair instability with $\langle \psi'_{i;\beta}\psi'_{i;\alpha}\rangle = \alpha \Delta  \delta_{\alpha\bar\beta} $. 
We get the pairing mean-field channel as 
\begin{equation}
\begin{aligned}
    P_Gc^\dagger_{\mathbf{k};\lambda}\gamma(\mathbf{k})_{\lambda\alpha}f_{i;\alpha}^{}P_G
    =& c^\dagger_{\mathbf{k};\lambda}\gamma(\mathbf{k})_{\lambda\alpha}
    \left(\sum_\beta s^\dagger_{i;\beta}\psi'_{i;\beta}\psi'_{i;\alpha}
    +\sum_{\beta\gamma(\beta<\gamma)} t^{}_{i;\alpha\beta\gamma} \psi'^\dagger_{i;\beta}\psi'^\dagger_{i;\gamma}\right)\\
    \sim &  c^\dagger_{\mathbf{k};\lambda}\gamma(\mathbf{k})_{\lambda\alpha}\left(\alpha \Delta s^\dagger_{i;\bar{\alpha}}
    +\Delta^*\sum_{\beta(\beta<\bar\beta)} \beta t^{}_{i;\alpha\beta\bar{\beta}}\right)~.
\end{aligned}
\end{equation}
In the above decoupling of $c^\dagger\gamma f$, there will be another term $\sim \langle c^\dagger\gamma s + t^\dagger\gamma c\rangle \psi'\psi'$.  
We take this term as small compared to the spin induced pairing Eq.~\eqref{eqn:mf_dwave} and omitted. 
Note that only 8 independent fermions of the 56 triplon $t$ fermions are coupled to the other fermions in the mean-field level. 
We therefore introduce a recombined triplon fermion as:
$t_{i;\alpha} = \frac{1}{2\sqrt{3}} \sum_{\beta} \beta t_{i;\alpha\beta\bar{\beta}}$.
We use the same notation $t$ here for the recombined triplon and distinguish it with the original triplon by the number of subscript. 
In terms of this new triplon, 
\begin{equation}\label{eqn:f_sc}
    P_Gf_{i;\alpha}^{}P_G = \alpha\Delta s^\dagger_{i;\bar{\alpha}}
    +\sqrt{3}\Delta^*t^{}_{i;\alpha}~,
\end{equation}
which recovers the result of Ref.~\cite{zhao_mixed_2025}.

The other possible mean-field channel is the hybridization between charge neutral $\psi'$ and charged particles $s$, $t$ and $c_1$, 
%$B_s=\langle s^\dagger_{i;\alpha}\psi'_{i;\alpha}\rangle$. 
\begin{equation}
\begin{aligned}
    \sum_{\alpha,\lambda}P_Gc^\dagger_{\mathbf{k};\lambda}\gamma(\mathbf{k})_{\lambda\alpha}f_{i;\alpha}^{}P_G
    =& \sum_{\alpha,\lambda}c^\dagger_{\mathbf{k};\lambda}\gamma(\mathbf{k})_{\lambda\alpha}
    (\sum_\beta s^\dagger_{i;\beta}\psi'_{i;\beta}\psi'_{i;\alpha}
    +\sum_{\beta\gamma} t^{}_{i;\alpha\beta\gamma} \psi'^\dagger_{i;\beta}\psi'^\dagger_{i;\gamma})\\
    \sim&\sum_{\alpha,\beta,\lambda}\langle c^\dagger_{\mathbf{k};\lambda}\gamma(\mathbf{k})_{\lambda\alpha}\psi'_{i;\alpha}\rangle s^\dagger_{i;{\beta}} \psi'^{}_{i;\beta}
     + \sum_{\alpha,\beta,\lambda}c^\dagger_{\mathbf{k};\lambda}\gamma(\mathbf{k})_{\lambda\alpha}\psi'_{i;\alpha}\langle s^\dagger_{i;{\beta}} \psi'^{}_{i;\beta}\rangle\\
     &+\frac{1}{\sqrt{3}}\sum_{\alpha,\beta,\lambda}\alpha\langle c^\dagger_{\mathbf{k};\lambda}\gamma(\mathbf{k})_{\lambda\alpha}\psi'^\dagger_{i;\bar\alpha}\rangle \psi'^\dagger_{i;\beta}t_{i;{\beta}} 
     +\frac{1}{\sqrt{3}}\sum_{\alpha,\beta,\lambda}\alpha c^\dagger_{\mathbf{k};\lambda}\gamma(\mathbf{k})_{\lambda\alpha}\psi'^\dagger_{i;\bar\alpha}\langle \psi'^\dagger_{i;\beta}t_{i;{\beta}} \rangle~,
\end{aligned}
\end{equation}
where we have assumed $\sum_\lambda \langle c^\dagger_{\mathbf{k};\lambda}\gamma(\mathbf k)_{\lambda\alpha}\psi'^\dagger_{i;\beta}\rangle\propto \alpha \delta_{\alpha,\bar\beta}$. 
This relies on a strong pairing $\Delta$ such that only the 8 tripons introduced above are important.

\begin{figure}[t]
    \centering
    \includegraphics[width=0.5\linewidth]{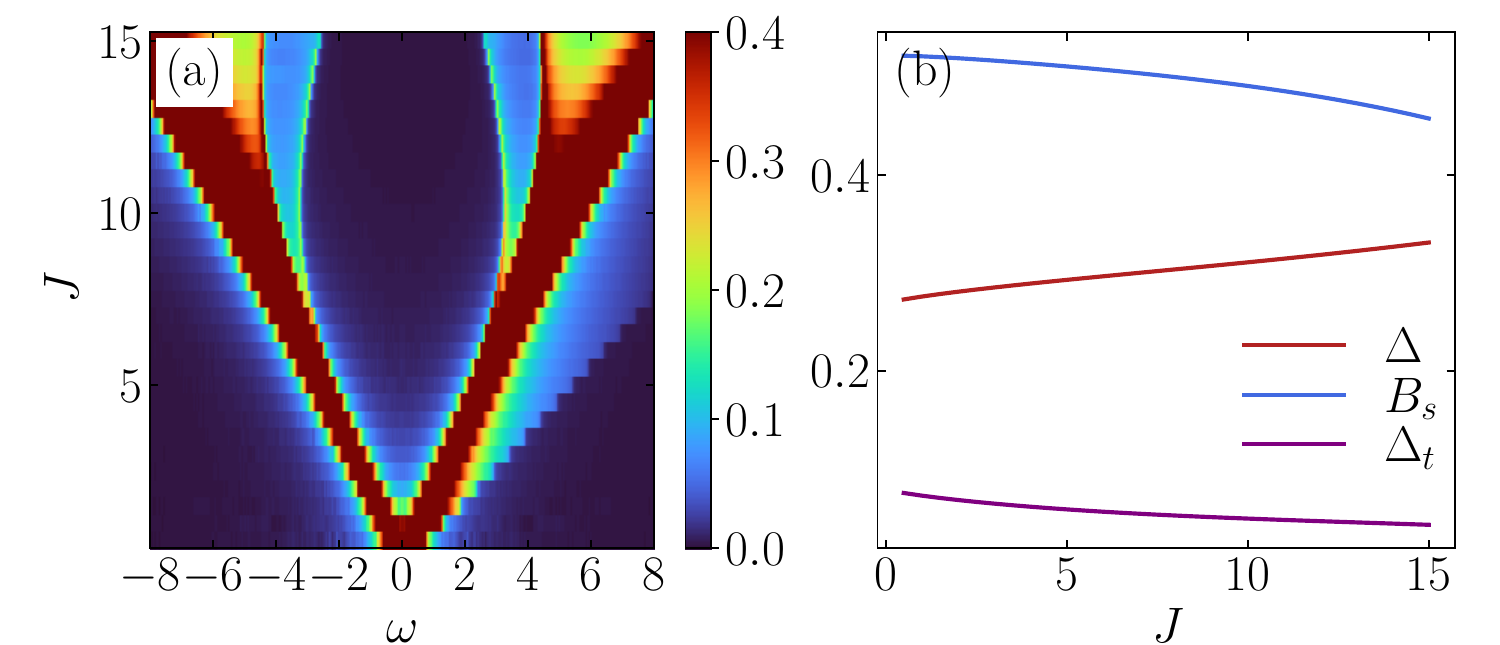}
    \caption{Superconducting evolution as a function of spin coupling $J$, with the coefficient $\alpha_B=1.0$ in Eq.~\eqref{eqn:self_consistent}. The parameters are set to $\theta=1.06^\circ, w_0/w_1=0.8$, $U=25$ meV, $M=0$ and $\nu=-2.4$. 
    (a) The STM spectra and (b) the corresponding order parameters for different value of $J$. 
    }
    \label{fig:stm_B7}
\end{figure}

We can now write down the full mean-field Hamiltonian as 
\begin{equation}\label{eqn:THFM_mf_app}
\begin{aligned}
    H_{\mathrm{MF}} = &H_0^{(c_1,c_2)} - \mu N_c\\
    &+ \sum_{\mathbf k,\mathbf G,\alpha\beta} c^\dagger_{1,\mathbf{k+G};\alpha}\gamma(\mathbf{k+G})_{\alpha\beta}(\beta \Delta s^\dagger_{-\mathbf k;\bar\beta}+\sqrt{3}\Delta^* t_{\mathbf k;\beta})+\mathrm{h.c.}
    - 2J\sum_{\mathbf{k},\alpha}\Delta^* \alpha \psi'_{-\mathbf{k};\bar\alpha}\psi'_{\mathbf{k};\alpha} + \mathrm{h.c.}\\
    &+ \sum_{\mathbf k,\mathbf G,\alpha\beta} c^\dagger_{1,\mathbf{k+G};\alpha}\gamma(\mathbf{k+G})_{\alpha\beta}(B_s \psi'_{\mathbf k;\beta} + \beta \Delta_t \psi'^\dagger_{-\mathbf {k};\bar\beta})+\mathrm{h.c.}
     +\sum_{\mathbf{k};\alpha} (B'_s s^\dagger_{\mathbf{k};\alpha}\psi'_{\mathbf{k};\alpha} + \Delta'_tt^\dagger_{\mathbf{k};\alpha} \psi'_{\mathbf{k};\alpha} )
    +\mathrm{h.c} \\
    & +  (E_s+\mu-\lambda) \sum_{i;\alpha} s^\dagger_{i;\alpha} s^{}_{i;\alpha} 
    -\frac{1}{2}\lambda\sum_{i;\alpha}\psi'^\dagger_{i;\alpha} \psi'_{i;\alpha} 
    + (E_t-\mu-\lambda)\sum_{i;\alpha}t^\dagger _{i;\alpha} t^{}_{i;\alpha} + \mathrm{const.}~,
\end{aligned}
\end{equation}
where the order parameters are solved iteratively as: 
\begin{equation}\label{eqn:self_consistent}
\begin{aligned}
    \Delta =& \frac{1}{N}\sum_{\mathbf k}\alpha\langle \psi'_{-\mathbf{k};\bar \alpha}\psi'_{\mathbf{k};\alpha}\rangle~,\\
    B_s =& 7\alpha_B\frac{1}{N}\sum_{\mathbf k} \langle s^\dagger_{\mathbf{k};\alpha} \psi'^{}_{\mathbf{k};\alpha}\rangle~,\\
    B'_s =& 7\alpha_B\frac{1}{N}\sum_{\mathbf{k};\alpha }\langle c^\dagger_{1;\mathbf{k};\alpha}\gamma_{\alpha\beta} \psi'^{}_{\mathbf{k};\beta}\rangle~,\\
    \Delta_t =& 2\sqrt{3}\alpha_B\frac{1}{N}\sum_{\mathbf k} \langle t^\dagger_{\mathbf{k};\alpha} \psi'^{}_{\mathbf{k};\alpha}\rangle~,\\
    \Delta'_t =& 2\sqrt{3}\alpha_B\frac{1}{N}\sum_{\mathbf{k};\alpha }\beta\langle c^\dagger_{1;\mathbf{k};\alpha}\gamma_{\alpha\beta} \psi'^\dagger_{\mathbf{k};\bar\beta}\rangle~,\\
\end{aligned}
\end{equation}
where $N$ is the total number of AA site in our calculation. 
The coefficient in front of $B_{s}, B'_s$ and $\Delta_t,\Delta_{t}'$ are determined by index counting. 
An additional phenomenological coefficient $\alpha_B <1$ is introduced to control the convergence of mean-field calculation. 
To validate an $\alpha_B$ smaller than the unity, we can consider the infinite $J$ limit, where the double occupied state of $f$ should be further projected to a subspace with lower spin energy. 
The coefficient will be much smaller in this limit. 
In practice, we consider a finite $J$ term and therefore take a finite $\alpha_B$.
Specifically, we set $\alpha_B = 0.5$, which leads to good numerical convergence.
In Fig.~\ref{fig:stm_B7}, we also present the STM spectra as well as the mean-field order parameters as a function of $J$ for $\alpha_B = 1.0$.
In this case, the condensation of the slave boson $B$ is overestimated, resulting in a superconductivity that is much stronger than in reality.
We note this is one limitation of mean-field calculation and leave further validation to variational Monte Carlo calculation calculation. 

Finally, in extracting the spectrum function at the mean-field level, we rewrite the $f$ fermion as: 
 
\begin{equation}
\begin{aligned}
    P_G f_{i;\alpha} P_G 
    =&\Delta \alpha s^\dagger_{i;\bar{\alpha}} +
    \sqrt{3}\Delta^* t^{}_{i;\alpha} 
    +B_s \psi'_{i;\alpha} + \sqrt{3}\alpha\Delta^*_t \psi'^\dagger_{i;\bar{\alpha}}~,
\end{aligned}
\end{equation}
%which necessarily host SC due to the hybridization of $s$ and $s^\dagger$. 
which is used in the Lehmann representation for the spectrum.
In reality, the mean-field parameters $B_s,\Delta,\Delta_t$ have their own fluctuations, and the real spectrum should be much broadened than our mean-field result.

\section{Band and gap structures of sFL and FL. }
\begin{figure}[t]
    \centering
    \includegraphics[width=0.95\linewidth]{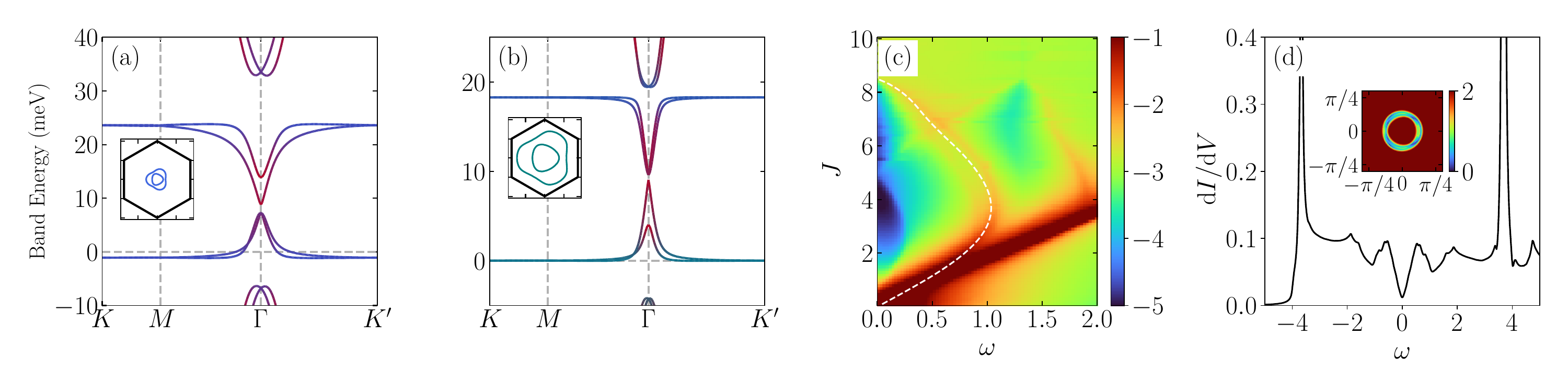}
    \caption{
    Band structures for the sFL and FL phases without invoking superconductivity.
    (a) Band structure of the sFL phase obtained by setting $B_s = B_s' = \Delta_t = \Delta_t' = 0$. The red and blue colors represent the spectral weight of the $c$ fermion and the $s,t$ fermions, respectively. 
    The inset shows the small Fermi surface dominated by the $s$ fermion.
    (b) Band structure of the FL phase obtained by setting $\Delta = \Delta_t = \Delta_t' = 0$. 
    The red, blue, and teal colors represent the spectral weight of the $c$, $s$, and $\psi'$ fermions, respectively. 
    The inset shows the large Fermi surface dominated by the $\psi'$ fermion.  
    (c) Fitting of the superconducting gap using Eq.~\ref{eqn:scgap} (dashed white line) for $x = 0.4$. The density of states is shown on a logarithmic scale to clearly reveal the lower-gap features. The same parameters as in Fig.~\ref{fig:stm_JA} are used.  
    (d) Spectrum obtained from the slave-rotor model [Eq.~\ref{eqn:rotormf}] with $B = 0.1$, $J \Delta = 3$ meV, and $\mu_\psi = -2.1$ meV. 
    The inset illustrates the momentum-space gap profile, showing a $d_{x^2-y^2}$-wave gap structure on a small Fermi surface dominated by the $c$ fermion.
    }
    \label{fig:bands}
\end{figure}

\subsection{sFL and FL as normal states. }

Here we present the band structures of the two different normal states corresponding to the sFL and FL phases, obtained without invoking superconductivity.

We first consider the strong-$J$ limit, where the pairing amplitude $\Delta \neq 0$ while the slave bosons are gapped, i.e., $B_s = B'_s = \Delta_t = \Delta'_t = 0$.
In this limit, the mean-field Hamiltonian in Eq.~\eqref{eqn:THFM_mf_app} simplifies to
\begin{equation}\label{eqn:sFL_mf}
\begin{aligned}
    H_{\mathrm{MF}}^{\mathrm{sFL}} = &H_0^{(c_1,c_2)}-\mu N_c 
    + \sum_{\mathbf k,\mathbf G,\alpha\beta} c^\dagger_{1,\mathbf{k+G};\alpha}\gamma(\mathbf{k+G})_{\alpha\beta}(\beta \Delta s^\dagger_{-\mathbf k;\bar\beta}+\sqrt{3}\Delta^* t_{\mathbf k;\beta})+\mathrm{h.c.}\\
    & +  (E_s+\mu-\lambda) \sum_{i;\alpha} s^\dagger_{i;\alpha} s^{}_{i;\alpha} 
    + (E_t-\mu-\lambda)\sum_{i;\alpha}t^\dagger _{i;\alpha} t^{}_{i;\alpha} \\
    &- 2J\sum_{\mathbf{k},\alpha}\Delta^* \alpha \psi'_{-\mathbf{k};\bar\alpha}\psi'_{\mathbf{k};\alpha} + \mathrm{h.c.} + 
    -\frac{1}{2}\lambda\sum_{i;\alpha}\psi'^\dagger_{i;\alpha} \psi'_{i;\alpha} 
    \mathrm{const.}~. 
\end{aligned}
\end{equation}
In this case, the $\psi'$ fermions form tightly bound local pairs and become decoupled from the conducting bands.
The resulting band structure of the $c$, $s$, and $t$ fermions is shown in Fig.~\ref{fig:bands}(a) for doping $x = 0.5$.
The original flat band of TBG splits into an upper Hubbard band dominated by the $t$ fermion and a lower Hubbard band dominated by $s^\dagger$.
Upon doping, the holes primarily occupy the lower Hubbard band near the $\Gamma$ point, 
where the Fermi surface is dominated by the $s^\dagger$ fermion.
The Fermi surface volume per spin and valley can be counted as $A_{\mathrm{sFL}} = (n_c + n_t-n_s)/4 = -x/4$, 
where we have used the relations $2 - x = n_c + n_s + n_{\psi'} + 3 n_t$ and $n_s + n_{\psi'}/2 + n_t = 1$.
This is consistent with the small Fermi surface shown in the inset of Fig.~\ref{fig:bands}(a).

In the opposite limit, where $J \rightarrow 0$ and the pairing order parameters are gapped ($\Delta = \Delta_t = \Delta_t' = 0$), the system reduces to a renormalized heavy-fermion model:
\begin{equation}\label{eqn:hFL_mf}
\begin{aligned}
    H_{\mathrm{MF}}^{\mathrm{FL}} = &H_0^{(c_1,c_2)}-\mu N_c
    + \sum_{\mathbf k,\mathbf G,\alpha\beta} c^\dagger_{1,\mathbf{k+G};\alpha}\gamma(\mathbf{k+G})_{\alpha\beta}B_s \psi'_{\mathbf k;\beta} +\mathrm{h.c.}
    +\sum_{\mathbf{k};\alpha} B'_s s^\dagger_{\mathbf{k};\alpha}\psi'_{\mathbf{k};\alpha} 
    +\mathrm{h.c} \\
    & +  (E_s-\mu-\tilde\lambda) \sum_{i;\alpha} s^\dagger_{i;\alpha} s^{}_{i;\alpha} 
    -(\mu+\frac{1}{2}\tilde\lambda)\sum_{i;\alpha}\psi'^\dagger_{i;\alpha} \psi'_{i;\alpha} 
    + \mathrm{const.}~,
\end{aligned}
\end{equation}
where the $t$ fermions becomes decoupled and omitted. 
We further redefine the Lagrange multiplier as $\lambda = \tilde{\lambda} + 2\mu$ so that both the $s$ and $\psi'$ fermions carry charge $+e$.
The resulting band structure of the $c$, $s$, and $\psi'$ fermions is shown in Fig.~\ref{fig:bands}(b), where different colors represent their respective spectral weights.
Unlike in the sFL phase, the low-energy states are now dominated by the $\psi'$ fermion, while the $s$ fermion forms the upper Hubbard band.
The Fermi surface volume per spin and valley can be counted as $A_{\mathrm{FL}} = (n_c + n_s+n_\psi')/4 = (2-x)/4$, 
where we have used the relation $2 - x = n_c + n_s + n_{\psi'} + 3n_t$ and $n_t = 0$.
This is consistent with the large Fermi surface shown in the inset of Fig.~\ref{fig:bands}(b).

\subsection{Superconducting gap size. }

Starting from the normal-state phases, sFL and FL, we can estimate the superconducting gap size based on the parent states.
Near the sFL phase, the slave-boson amplitudes $B_s$ and $B_s'$ are small.
By integrating out the $c_1$, $c_2$, $t$, and $\psi'$ fermions, we obtain an effective Hamiltonian for the $s$ fermion, in which the pairing term scales as $\sim \frac{B_s^2}{J}  \alpha s^\dagger_{-\mathbf{k};\bar{\alpha}} s^\dagger_{\mathbf{k};\alpha}$.
In contrast, near the FL phase, integrating out the $c_1$, $c_2$, and $s$ fermions yields an effective Hamiltonian for the $\psi'$ fermion.
Here, the $\psi'$ fermions are directly paired by the term $-2J\Delta \alpha \psi'^\dagger_{-\mathbf{k};\bar{\alpha}} \psi'^\dagger_{\mathbf{k};\alpha}$.

To interpolate between these two limits, we propose the following phenomenological form for the superconducting gap:
\begin{equation}\label{eqn:scgap}
    E_{\mathrm{gap}} = \frac{a|B_s|'^2J}{|B_s'|^2+J^2|b\Delta|^2}, 
\end{equation}
where $a$ and $b$ are fitting parameters. 
In the sFL limit ($B_s' \ll J$), the gap scales as $E_{\mathrm{gap}} \sim |B_s'|^2 / J$, while in the FL limit ($B_s' \gg J$), it approaches $E_{\mathrm{gap}} \sim J$.
In Fig.~\ref{fig:bands}(c), we find the gap function Eq.~\eqref{eqn:scgap} shows an overall good agreement with the actual mean-field result, 
where $a = 4.5$ and $b = 32$ is used in our fitting.

In both the sFL and FL phases, the superconducting state exhibits a nematic $p_x$-wave–like structure, with a pair of nodes appearing along the $k_y$ direction for each of the two Fermi surfaces.
When the superconducting gap $E_{\mathrm{gap}}$ becomes sufficiently large, the nodes from the two Fermi surfaces merge, resulting in a fully gapped spectrum.
The gap function in Eq.~\eqref{eqn:scgap} thus suggests that the nodal gap structure is more robust in the sFL phase,
where the superconducting gap increases quadratically as $B_s^2$ and can persist under stronger slave-boson condensation.
Superconductivity induced by on-site spin pairing in the FL phase has also been discussed in Ref.~\cite{wang2024molecular},
where a gapless nodal gap structure appears only in the very weak-coupling limit of small $J$.
This behavior is consistent with our results in Fig.~\ref{fig:bands}(c),
where a fully gapped spectrum emerges for $J \sim 1$–$5$ meV, closer to the FL side of the phase diagram.

\subsection{Slave rotor description of FL phase.}

The slave rotor model is a commonly used slave-particle approach to describe heavy-fermion systems \cite{Lau2023THFM}. 
In this model, the localized $f$ fermion is fractionalized as $f_{i;\alpha} = B \psi_{i;\alpha}$,
where $B = e^{i\theta}$ is a bosonic rotor carrying the charge, while $\psi_{i;\alpha}$ is a fermionic spinon carrying the flavor.

Below we briefly discuss the possible phases of the slave rotor model.
We start from the mean-field Hamiltonian,
\begin{equation}\label{eqn:rotormf}
\begin{aligned}
    H_{\mathrm{MF}}^\mathrm{rotor} = &H_0^{(c_1,c_2)}-\mu N_c + H_{\mathrm{MF}}^{(B)}[B]
    + \sum_{\mathbf k,\mathbf G,\alpha\beta} c^\dagger_{1,\mathbf{k+G};\alpha}\gamma(\mathbf{k+G})_{\alpha\beta}B\psi_{\mathbf{k};\beta}+\mathrm{h.c.}
    +  \\
    &- 2J\sum_{\mathbf{k},\alpha}\Delta^* \alpha \psi_{-\mathbf{k};\bar\alpha}\psi_{\mathbf{k};\alpha} + \mathrm{h.c.}
    -(\mu+\mu_\psi)\sum_{i;\alpha}\psi^\dagger_{i;\alpha} \psi_{i;\alpha} + \mathrm{const.}, 
\end{aligned}
\end{equation}
where we adopt the same mean-field pairing Hamiltonian for the $\psi$ fermion as that of $\psi'$ in Eq.~\eqref{eqn:THFM_mf_app}.
Here, $\mu_\psi$ denotes the chemical potential shift imposed by the particle-number constraint,
and $H_{\mathrm{MF}}^{(B)}$ represents the mean-field Hamiltonian of the rotor sector.

We do not attempt to solve the self-consistent mean-field equations here, but only outline the possible phases.
When $B\gamma \gg J\Delta$, the Fermi surface is dominated by the $\psi$ fermion.
In this regime, the slave rotor model reproduces the same physics as the Fermi liquid phase discussed in Eq.~\eqref{eqn:hFL_mf} in the small $J$ limit,
featuring a single superconducting gap where the superconducting and RVB gaps are indistinguishable.

Conversely, when $B\gamma \ll J\Delta$, the $\psi$ fermion becomes gapped due to the spin exchange $J$,
and the Fermi surface is instead dominated by the $c$ fermion.
Since the $c$ fermion also forms a small Fermi liquid, this regime shares some similarity to sFL phase.
In Fig.~\ref{fig:bands}(d), we illustrate a possible superconducting state in this scenario using a representative parameter set without solving the mean-field equations.
A two-gap structure is also observed: the upper gap $\sim 4$ meV corresponds to the RVB gap induced by $\psi$,
while the lower V-shaped gap $\sim 0.7$ meV originates from the superconductivity mainly associated with the $c$ fermion.
The feature near $2$ meV corresponds to an interband gap between the two Fermi surfaces.
However, this V-shaped gap appears only when the active bandwidth $M$ is sufficiently large,
and the pairing symmetry in this case is $d_{x^2 - y^2}$ rather than the $p_x$-wave–like nematic pairing discussed in our scenario.

\end{widetext}

\section{More mean-field results}\label{app:moreMF}

This appendix presents supplementary mean-field results, including analyses of alternative pairing channels and variations in model parameters, such as the Hubbard interaction and finite strain.

\subsection{$s$-wave pairing channel}

\begin{figure}[t]
    \centering
    \includegraphics[width=0.95\linewidth]{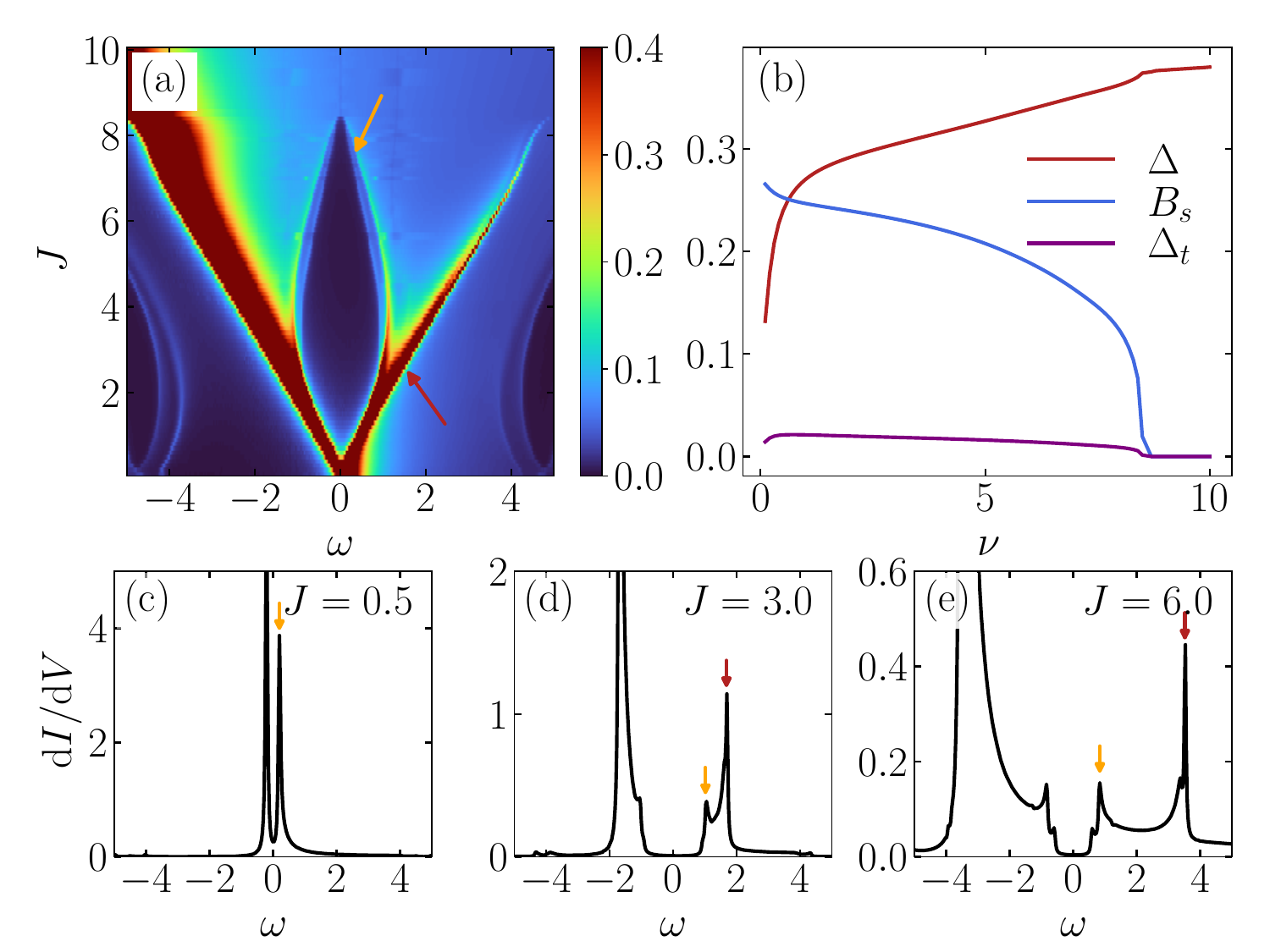}
    \caption{Superconducting evolution as a function of spin coupling $J$ when $s$-wave pairing is considered, for $\theta=1.06^\circ, w_0/w_1=0.8$, $U=25$ meV, $M=0$ and $\nu=-2.4$. 
    (a) The STM spectrum and (b) the order parameters for different value of $J$. 
    (c) (d) and (e) show three different line cuts of STM at $\nu=-2.9$, $\nu=-2.6$ and $\nu=-2.3$, respectively. 
    %(c) The STM spectrum and (d) the order parameters for different value of $J$ for a fixed doping $\nu=-2.4$.  
    }
    \label{fig:stm_swave_JA}
\end{figure}

\begin{figure}[t]
    \centering
    \includegraphics[width=0.95\linewidth]{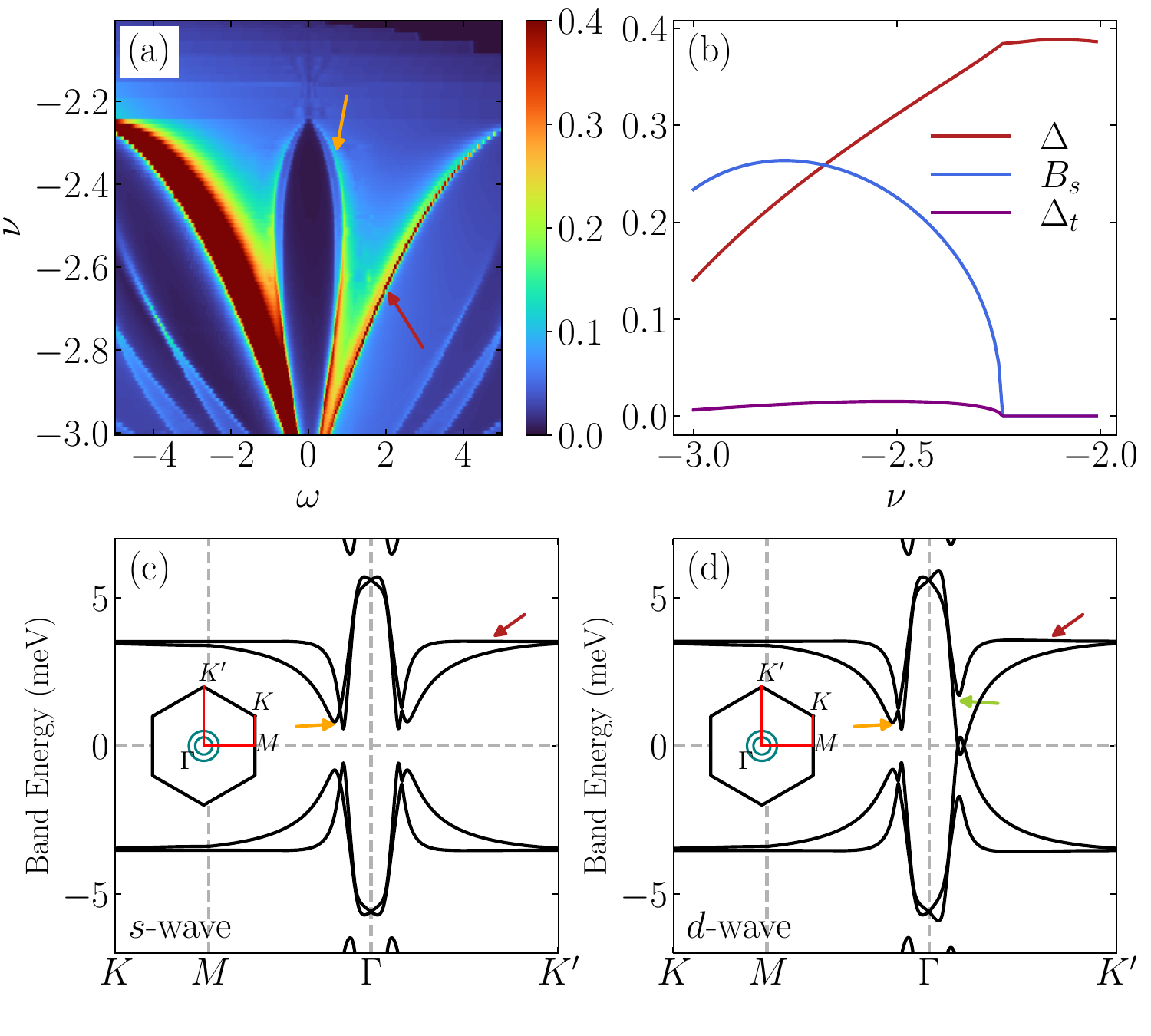}
    \caption{Superconducting evolution as a function of charge density $\nu$ when $s$-wave pairing is considered, for $\theta=1.06^\circ, w_0/w_1=0.8$, $U=25$ meV $J=6$ meV and $M=0$. 
    (a) The STM spectrum and (b) the order parameters for different value of $\nu$. 
    (c) and (d) show the BdG bands for both (c) $s$-wave and (d) $d$-wave pairing channel at the same parameter $x=0.4$ and $J=6$ meV. 
    The red, orange, and green arrows indicate the pseudogap, superconducting gap, and inter-band gap, respectively.
    The inset shows the normal-state Fermi surface and the momentum cut used in (c) and (d).
    }
    \label{fig:stm_swave}
\end{figure}

In the main text, we focus on the $d$-wave pairing states.
However, $s$-wave pairing is also allowed within certain parameter regimes.
Figs.~\ref{fig:stm_swave_JA} and \ref{fig:stm_swave} present the mean-field results at different doping levels $x$ and exchange couplings $J$ when the $s$-wave pairing channel is considered.
The high-energy features are essentially the same for both the $s$- and $d$-wave pairing channels.
A two-gap structure is also observed in the $s$-wave case, and the evolution of the order parameters as well as the pseudogap magnitude closely follow those of the $d$-wave case.
We therefore conclude that the RVB pairing mechanism based on an sFL provides a general framework that can accommodate multiple pairing symmetries.

The distinction between $s$- and $d$-wave pairing symmetries arises mainly from the low-energy behavior.
At low energies, a characteristic $U$-shaped superconducting gap appears for all doping levels $x$ and exchange couplings $J$, consistent with the $s$-wave pairing symmetry.
Additionally, the inter-band gap present in the $d$-wave case [cf. green arrows in Figs.~\ref{fig:stm_JA} and \ref{fig:stm_noproj}] is absent for the $s$-wave pairing channel, as shown in the spectra in Figs.~\ref{fig:stm_JA}(c) and (d).
In other words, the inter-band gap is closed in the $s$-wave case.

A simple understanding can be obtained as follows.
Let $\tilde \psi_{1\eta s}$ and $\tilde \psi_{2\eta s}$ denote the two active bands in the band index, and $\psi_+$ and $\psi_-$ denote the same band in orbital index with $L_z = \pm 1$.
As discussed later in App.~\ref{app:nodes} 1., the two orbitals $\psi_\pm$ are split by a mass term of the form $m_x\sigma_x+m_y\sigma_y$, which gives 
\begin{equation}\label{eqn:psiact}
\begin{aligned}
    \tilde \psi_{\mathbf{k};1\eta s} =& e^{i \phi(\eta\mathbf{k})/2}\psi_{\mathbf{k};+\eta s} + e^{-i \phi(\eta\mathbf{k})/2}\psi_{\mathbf{k};-\eta s}~, \\
    \tilde \psi_{\mathbf{k};2\eta s} =& e^{i \phi(\eta\mathbf{k})/2}\psi_{\mathbf{k};+\eta s} - e^{-i \phi(\eta\mathbf{k})/2}\psi_{\mathbf{k};-\eta s}~. 
\end{aligned}
\end{equation}
For the $s$-wave pairing, the pairing occurs between different orbitals, $\psi_+$ and $\psi_-$, such that
$\langle \tilde \psi_{-\mathbf{k};2\bar \eta \bar s}\tilde\psi_{\mathbf{k};1 \eta s}\rangle \sim
\langle \psi_{-\mathbf{k};+\bar\eta\bar s}\psi_{\mathbf{k};-\eta s}\rangle-\langle\psi_{-\mathbf{k};-\bar\eta\bar s}\psi_{\mathbf{k};+\eta s}\rangle =0$. 
In contrast, for the $d$-wave pairing, the pairing occurs between the same orbital components, $\psi_+\psi_+$ or $\psi_-\psi_-$, leading to
$\langle \psi_{-\mathbf{k};2\bar\eta\bar s}\psi_{\mathbf{k};1\eta s}\rangle \sim\langle \psi_{-\mathbf{k};+\bar\eta\bar s}\psi_{\mathbf{k};+\eta s}\rangle e^{i\phi(\eta\mathbf{k})}-\langle \psi_{-\mathbf{k};-\bar\eta\bar s}\psi_{\mathbf{k};-\eta s}\rangle e^{-i\phi(\eta\mathbf{k})} \neq 0$. 
The latter expression remains finite when the phase $\phi(\mathbf{k})$ is nontrivial, resulting in a nodal structure of the inter-band pairing gaps in the $d$-wave pairing case.

\subsection{Intra-valley pairing}

\begin{figure}[t]
    \centering
    \includegraphics[width=0.95\linewidth]{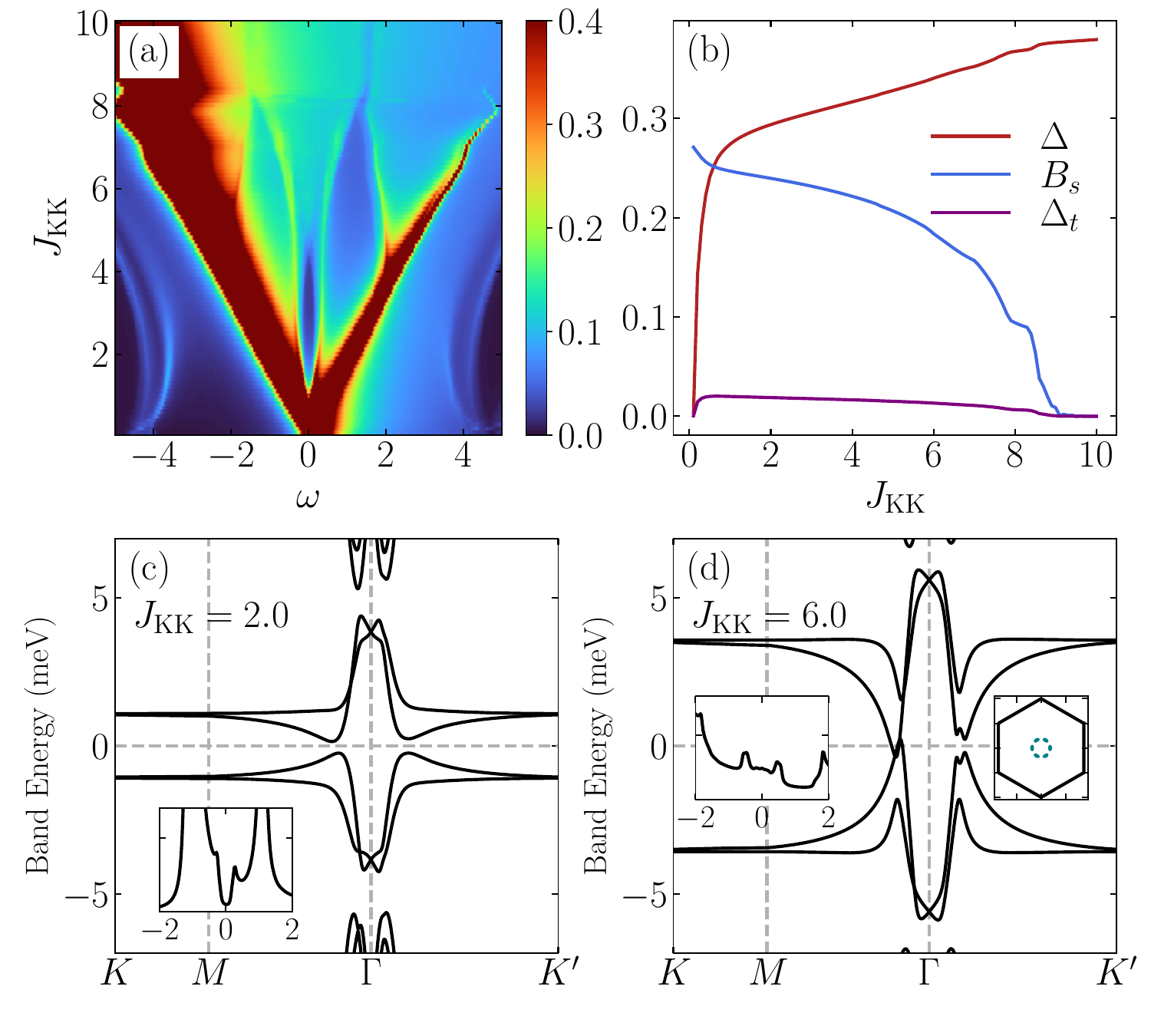}
    \caption{Superconducting evolution as a function of $J$ when an intra-valley $s$-wave pairing is considered, for $\theta=1.06^\circ, w_0/w_1=0.8$, $x=0.4$, $U=25$ meV. 
    A small mass $M=2.53$ meV is included here to avoid perfect nesting for intra-valley pairing. 
    (a) The STM spectrum and (b) the order parameters for different value of $J$. 
    (c) shows the typical BdG bands for interchange coupling $J_{\mathrm{KK}}=2$ meV. 
    The corresponding STM spectrum are shown in the inset, where a $U$-shape spectrum is seen. 
    (d) shows the BdG bands for interchange coupling $J_{\mathrm{KK}}=6$ meV. 
    The corresponding STM spectrum and the Bogoliubov Fermi surface are shown in the left and right insets, respectively. 
    An gapless SC with non vanishing zero bias density of states is seen. 
    The line cuts in (c) and (d) are chosen the same as those in Figs.~\ref{fig:stm_swave} (c) and (d). } 
    \label{fig:stm_KK}
\end{figure}

Although we attribute the pairing instabilities primarily to the inter-valley $d$-wave or $s$-wave channels, as discussed in App.~\ref{app:THFM}, 
we emphasize that our formalism can be readily generalized to other possible pairing channels. 
In this appendix, we explore the possibility of intra-valley spin-singlet pairing between flavors $\alpha=+K\uparrow$ and $\alpha'=-K\downarrow$. 
This pairing channel was also considered in Ref.~\cite{wang_kekule_2025}, where a Bogoliubov Fermi surface appears for small pairing strengths, giving rise to a $V$-shaped gap structure.

In Fig.~\ref{fig:stm_KK}, we show the mean-field results for an intra-valley $s$-wave RVB pairing described by the spin Hamiltonian:
\begin{equation}
    H_{J,\mathrm{MF}}^{s,KK} = -2J_{KK} \sum_{i;a\eta s} \Delta_{KK}^* s \psi'_{i;\bar a\eta\bar s} \psi'_{a\eta s}+\mathrm{h.c.}~,
\end{equation}
where we use subscript $KK$ to denote the intra-valley pairing. 
The order parameters evolve similarly to those of the inter-layer $d$-wave [c.f. Fig.~\ref{fig:stm_JA}] and $s$-wave [c.f. Fig.~\ref{fig:stm_swave_JA}] pairings.

However, the superconducting gap behavior is very different from the inter-valley pairing. 
For sufficiently large $J_{KK}$, the system remains in the sFL phase, and no superconductivity is transferred to the conduction band. 
As $J_{\mathrm{KK}}$ decreases, the slave boson $B$ condenses. 
Unlike the inter-valley pairing, the intra-valley pairing does not exhibit perfect nesting of the Fermi surface between $\mathbf{k}$ and $-\mathbf{k}$, so the Fermi surface is not fully gapped. 
Consequently, a Bogoliubov Fermi surface coexists with superconductivity in the weak pairing limit, leading to a STM spectrum with finite zero-bias density of states. 
When $J_{\mathrm{KK}}$ is further reduced and $B_s$ becomes larger, the superconductivity strengthens sufficiently to fully gap the Bogoliubov Fermi surface, resulting in a conventional $U$-shaped gap.

Although a gapless superconducting state is also possible in this pairing channel, we do not observe the typical $V$-shaped gap reported in recent experiments.
We emphasize that the central idea of our theory is to regard the sFL as the parent state, which naturally accounts for both the two-gap superconducting structure and the pseudogap normal state.
The intra-valley pairing results demonstrate that our framework is robust against various pairing channels,
and other possibilities involving symmetry-breaking orders can also be explored in future studies.

\subsection{Effect of Hubbard interaction $U$. }

\begin{figure}[t]
    \centering
    \includegraphics[width=0.95\linewidth]{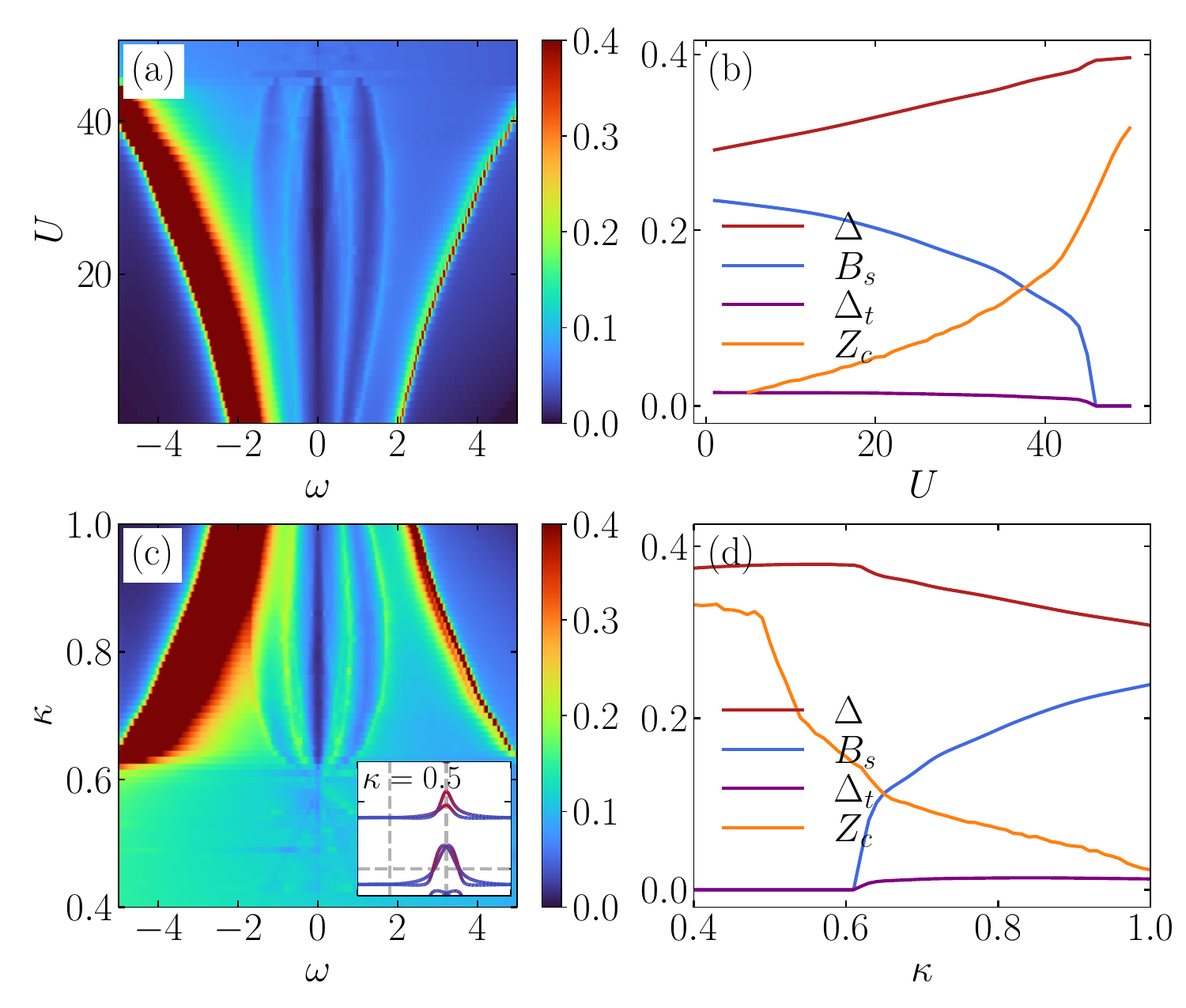}
    \caption{
    Superconducting evolution as the Hubbard interaction varies. 
    Here the $d$-wave pairing channel is considered, with parameters  $\theta = 1.06^\circ$, $w_0/w_1 = 0.8$, $x = 0.4$, $J = 6~\mathrm{meV}$, and $M = 0$. 
    (a) STM spectra and (b) order parameters for different values of $U$ at fixed $\kappa = 0.8$. 
    The spectral weight of the $c$ electron, $Z_c$, on the Fermi surface of the normal state ($B_s = 0$) is also shown. 
    (c) STM spectra and (d) order parameters for different values of $\kappa$ at fixed $U = 25~\mathrm{meV}$. 
    The spectral weight $Z_c$ on the normal-state Fermi surface is again displayed. 
    The inset of (c) shows the normal-state sFL band structure ($B_s = 0$) for $\kappa = 0.5, U=25~\mathrm{meV}$.
    }
    \label{fig:stm_U}
\end{figure}

In Fig.~\ref{fig:stm_U}, we present mean-field results obtained by varying both the Hubbard interaction $U$ and the phenomenological coefficient $\kappa$ in Eq.~\eqref{eqn:interactionf},  
at fixed doping $x=0.4$ and exchange coupling  $J=6.0$ meV. 
Increasing $U$ further stabilizes the pseudogap by enhancing $\Delta$ while suppressing $B_s$. 
For sufficiently large $U$, the superconducting gap eventually disappears even though the system remains in the sFL phase.
This occurs because, in the large-$U$ limit, the local RVB singlet on the $f$ orbital becomes decoupled from the itinerant $c$ electrons.
This picture is supported by the increase of the  $c$-electron spectral weight $Z_c$ in the normal state, which accompanies the disappearance of superconductivity, as shown in Fig.~\ref{fig:stm_U}(b).
The system thus enters a trivial Kondo-breakdown phase, where the small Fermi surface is dominated by weakly interacting $c$ electrons and superconductivity does not develop.

A similar evolution occurs when $\kappa$ is varied.
For smaller $\kappa$, a larger fraction of the doped carriers occupies the  $c$ bands rather than the $f$ bands, as indicated by the normal-state band structure in the inset of Fig.~\ref{fig:stm_U}(c).
When $\kappa$ becomes sufficiently small such that the spectral weight of the $c$ electrons increases significantly, superconductivity is again suppressed and eventually destroyed.

\subsection{Effect of finite strain. }

\begin{figure}
    \centering
    \includegraphics[width=0.95\linewidth]{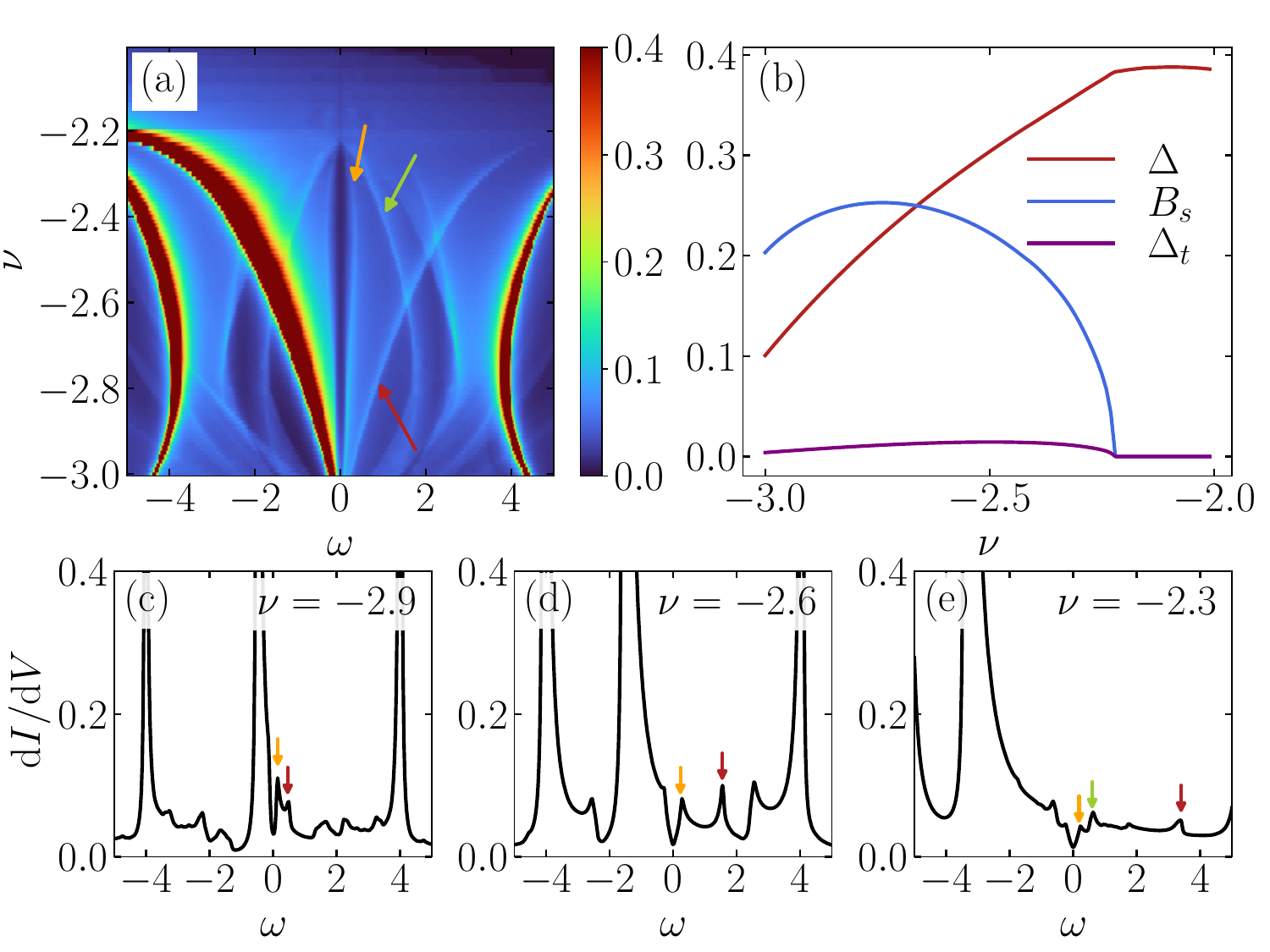}
    \caption{Superconducting evolution as a function of charge density $\nu$ with a finite strain $\epsilon_{xy}=3$ meV, for twist angle $\theta=1.06^\circ$, $w_0/w_1=0.8$, $U=25$ meV, $J=6$ meV and $M=0$. 
    (a) The STM spectrum and (b) the order parameters for different value of $\nu$.  
    (c) (d) and (e) show three different line cuts of STM at $\nu=-2.9$, $\nu=-2.6$ and $\nu=-2.3$, respectively. 
    }
    \label{fig:stm_strain}
\end{figure}

The ubiquitous effects of strain and lattice relaxation are important in TBG.  
In particular, Refs.~\cite{wagner2022global,kwan_kekule_2021} have shown that within Hartree–Fock theory, the ground state evolves from a K-IVC to an IKS state when finite strain is included.
Since the typical strain strength is considered comparable to the pseudogap energy scale, 
it is important to examine the stability of our results under finite strain.

To model the effect of heterostrain, we introduce a term that explicitly breaks lattice rotational symmetry,

\begin{equation}\label{eqn:strain}
    H_{\epsilon} = \epsilon_{xy}\sum_{i,\eta,s} f_{i;\eta+s}^\dagger f^{}_{i;\eta-s}+\mathrm{h.c.}~,
\end{equation}
whose effect is treated exactly in the $s$–$t$ subspace within our mean-field calculation.
Its influence on the $\psi'$ channel is neglected under the assumption that the $J$-term physics dominates.

Figure~\ref{fig:stm_strain} presents the results for $\epsilon_{xy}=3$ meV.
The evolution of the order parameters remains nearly unchanged from the unstrained case.
Both the superconducting gap and the pseudogap are suppressed by finite strain;
the pseudogap, however, is more robust, while the superconducting gap becomes smaller but remains non-zero.
A more detailed analysis shows that the superconducting gap is no longer nodal and acquires a small finite value $E_{\mathrm{gap}}\sim 0.01$ meV.
This minor gap is not resolvable in the STM spectrum and does not alter the characteristic $V$-shape.

\section{Analysis of the gap structure} \label{app:nodes}

\begin{figure}
    \centering
    \includegraphics[width=0.95\linewidth]{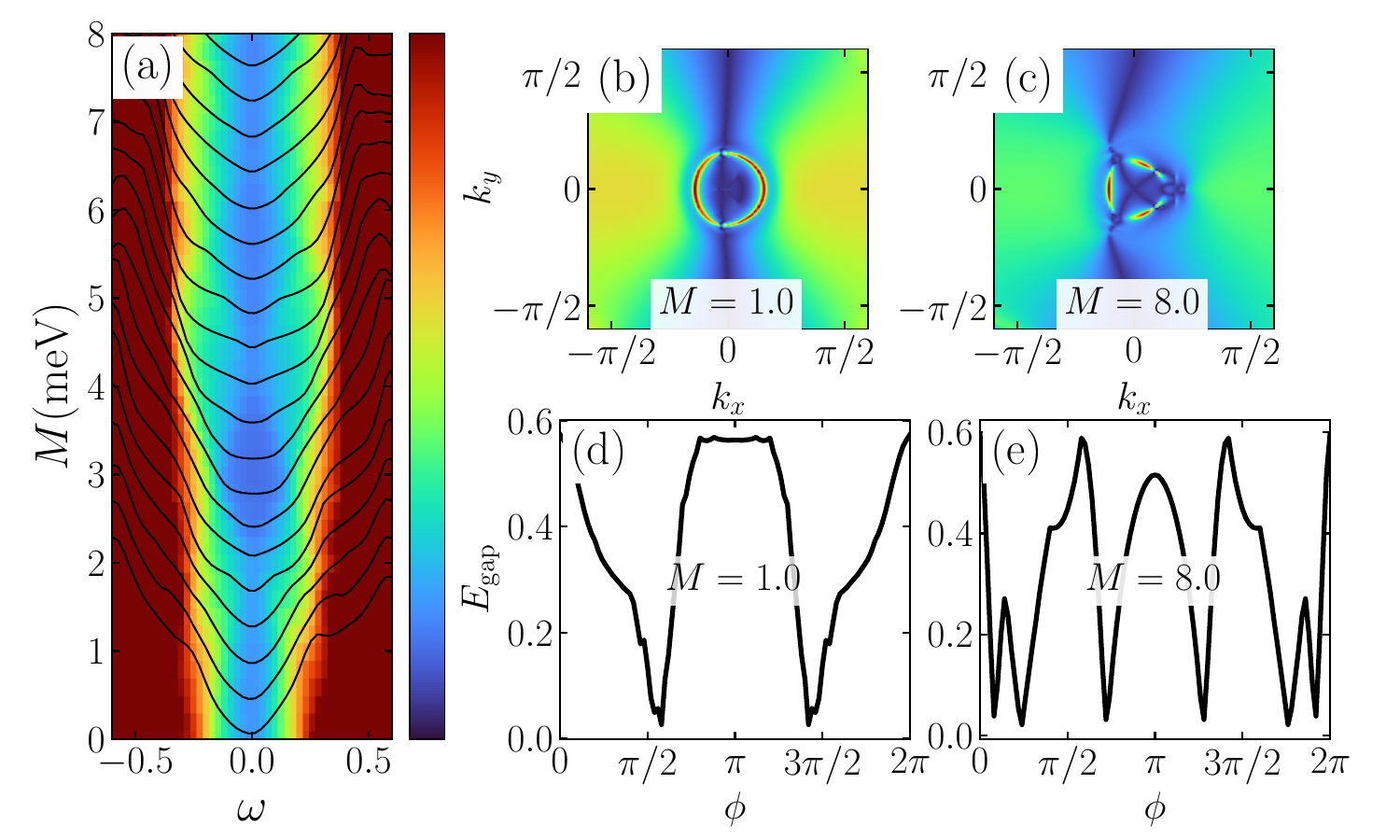}
    \caption{ Effect of the active band width $M$ on the pairing symmetry. 
    (a) STM spectrum as a function of $M$. 
    For small and large $M$, the STM show a V shape SC gap with node. 
    However, for intermediate value of $M\approx 3$ meV, there is a small gap in the spectrum. 
    (b) and (c) show the gap function $|\Delta_{\mathrm{act};1}(\mathbf{k})|$ in the momentum space projected into one of the lower Hubbard band for $M=1.0$ meV and $M=8.0$ meV, respectively. 
    (d) and (e) shows the minimal single particle gap along different azimuth angle $\phi$ for $M=1.0$ meV and $M = 8.0$ meV. 
    }
    \label{fig:pandd}
\end{figure}

\begin{figure}
    \centering
    \includegraphics[width=0.95\linewidth]{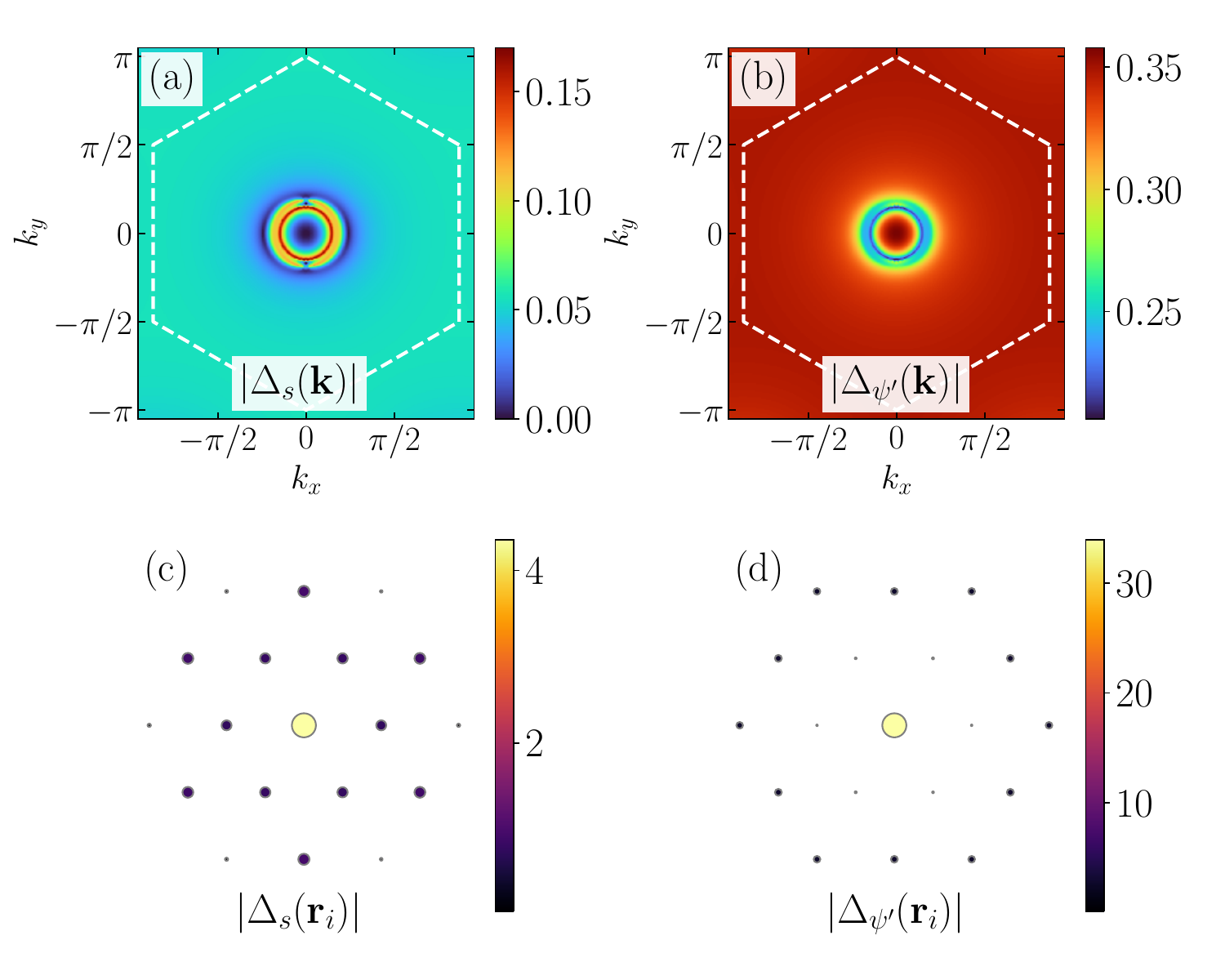}
    \caption{The pairing order parameter (a) $\Delta_s(\mathbf{k})=\alpha\langle s_{-\mathbf{k};\bar\alpha}s_{\mathbf{k};\alpha}\rangle$  for $s$ fermion and 
    (b) $\Delta_{\psi'}(\mathbf{k})=\alpha\langle \psi'_{-\mathbf{k};\bar\alpha}\psi'_{\mathbf{k};\alpha}\rangle$  for $\psi'$ fermion, calculated at twist angle  $\theta=1.06^\circ, w_0/w_1=0.8$, $U=25$meV, $J=6$ meV and doping $x=0.4$. 
    %(a) and (b) momentum space distribution, note the different colorbar for the two figures;
    Fourier transformed real space pairing for the (c) $s$ fermion 
    $\Delta_s(\mathbf{r}_i) = \frac{1}{\sqrt{N}}\sum_{\mathbf{k}} \Delta_s(\mathbf{k})e^{i\mathbf{k}\cdot \mathbf{r}_i}$
    and (d) $\psi'$ fermion 
    $\Delta_{\psi'}(\mathbf{r}_i) = \frac{1}{\sqrt{N}}\sum_{\mathbf{k}} \Delta_{\psi'}(\mathbf{k})e^{i\mathbf{k}\cdot \mathbf{r}_i}$.  Each circle represent an AA site of TBG.  
    Here the $s$ pair spread in the real space but $\psi'$ pair is localized on a single AA site.
    }
    \label{fig:deltak}
\end{figure}

\subsection{Origin of the nematic nodal structure.}

In this appendix, we give an analytical analysis of the nodal structure of the superconductor, and its dependence on the active band width $M$.  We use the d-wave pairing ansatz for $\psi'$. Note that this pairing is still purely on-site and has no momentum dependence.  In the SC phase with $B=0$, the mobile carriers dominated by $s^\dagger$ inherit the pairing, so we expect on-site pairing term $\Delta ( s_{i;+} s_{i;+}+s_{i;-}s_{i;-})$. Here we suppress the valley and spin index.  The pairing has no momentum dependence in the $(s_+(\mathbf k),s_-(\mathbf k))$ basis. However, we have two split Fermi surfaces due to the hybridization between $s_+$ and $s_-$, so the projected pairing to each split band acquires a momentum dependence. Therefore, it is important to understand the hybridization between $s_+, s_-$ and the structure of the resulting Bloch wavefunction due to the hybridization.  

We expect that the kinetic term of $s$ just follows that of $f$. So in the following, we try to integrate $c_1,c_2$ to get an effective single-particle Hamiltonian for $f$. We focus on the regime where the Fermi surface momentum $k_F$ is away from the $\Gamma$ point such that the $c_1,c_2$ band energy $v_*|\mathbf{k}|$ is the largest energy scale. 
An estimation based on the current parameter $\gamma=-26.184$ meV and $U=25$ meV shows that this condition is satisfied, except for the center $4.4\%$ area of the mini Brillouin zone, corresponding to the doping level $x\sim 0.35$. 
Therefore, we can integrate out the $c_1,c_2$ fermion in Eq.~\eqref{eqn:THFM_0} for most of the regime where superconducting emerges. 
After integration, the $f$ fermion gets an effective kinetic term: 
\begin{equation}
    H_{\mathrm{eff}}^{(f)} = \sum_{\mathbf{k}}  f^\dagger_{\mathbf k} h_{\mathrm{kin}}(\mathbf k) f_{\mathbf k}^{} + H_{\mathrm{int}}^{(f)}~, \\
\end{equation}
where 
\begin{equation}
\begin{aligned}
    h_{\mathrm{kin}}(\mathbf{k}) \approx %& H[f,c] \frac{1}{\mu-H[c_1,c_2]}H[c,f]\\
    & \frac{1}{v^2|\mathbf{k}|^2} \gamma^\dagger(\mathbf{k}) \left[
    \begin{matrix}
        -\mu & Mk^2/|\mathbf{k}|^2\\ M\bar{k}^2/|\mathbf{k}|^2& -\mu
    \end{matrix}\right]\tau_0 s_0 \gamma(\mathbf{k}) \\
    = & \left[\begin{matrix}
        \epsilon_{\mathrm{kin}}(\mathbf{k})& \delta_{\mathrm{kin}}(\mathbf{k})\\ \delta^*_{\mathrm{kin}}(\mathbf{k}) &\epsilon_{\mathrm{kin}}(\mathbf{k})
    \end{matrix}\right]\tau_0 s_0~,
\end{aligned}
\end{equation}
and 
\begin{equation}\label{eqn:projH_diag}
    \epsilon_{\mathrm{kin}}(\mathbf{k}) = \frac{e^{-|\mathbf{k}|^2\lambda^2}}{v^2|\mathbf{k}|^2} (-\mu\gamma_0^2-\mu v'^2|\mathbf{k}|^2)~,% + 2\mathrm{Re}(Mv'k^3/|k|^2))
\end{equation}
\begin{equation}\label{eqn:projH_offdiag}
    \delta_{\mathrm{kin}}(\mathbf{k}) = \frac{e^{-|\mathbf{k}|^2\lambda^2}}{v^2|\mathbf{k}|^2} 
    (\gamma_0^2Mk^2/|\mathbf{k}|^2-2\mu \gamma_0 v'\eta \bar{k})~,
    %+ (v'\bar{k})^2M\bar{k}^2/|k|^2)
\end{equation}
where we use the complex variable $k=k_x+ik_y$ and $\bar k = k_x - i k_y$. 
Here we keep only the most relevant first two terms for $\epsilon_{\mathrm{kin}}(\mathbf{k})$ and $\delta_{\mathrm{kin}}(\mathbf{k})$.  
The diagonal term $\epsilon_{\mathrm{kin}}(\mathbf{k})$ is the dispersion generated by hybridizing with $c_1,c_2$, 
which disappears at the charge neutrality point $\mu=0$. 
The off-diagonal term $\delta_{\mathrm{kin}}$ is the effective mass term which splits the two orbitals. 
Due to $C_{2z}T$ symmetry, the $\sigma_z$ mass term is forbidden. 

Note that $h_{\mathrm{kin}}$ simply gives an effective dispersion for $f$ orbital. Now we need to include the interaction $H^{(f)}_{\mathrm{int}}$. 
In the doped Mott insulator at filling $\nu = -2 - x$, our interest lies in the lower Hubbard band, 
which is  dominated by singlon $s$ fermion when $x$ is relatively large.  Then the $s$ fermion just inherits the effective kinetic term of the $f$ orbital.
The chemical potential now changes to around $\mu \sim -U/2$ due to the Hubbard interaction. 
Under the interaction, the diagonal term $\epsilon_{\mathrm{kin}}(\mathbf{k})$ in Eq.~\eqref{eqn:projH_diag} now effectively raises the energy near the $\Gamma$ point, producing an upward convex dispersion in the lower band near the $\Gamma$ point.
This is consistent with the band structure shown in the right inset of Fig.~\ref{fig:illu} (a). 

For the band-splitting mass term $\delta_{\mathrm{kin}}(\mathbf{k})$ in Eq.~\eqref{eqn:projH_offdiag}, the two distinct terms exhibit qualitatively different angular dependencies.
The first contribution, characterized by the active bandwidth $M$, 
produces a vortex structure of the Bloch wavefunction which winds twice, $\sim e^{2i\phi(\mathbf{k})}$, as ${\bf k}$ encircles around each of the two Fermi surfaces, 
where $\phi(\mathbf{k})$ is the polar angle of $\mathbf{k}$ measured from the $x$ axis.
This would lead to a superconductor with two nodal lines at each Fermi surface. 

The second contribution is proportional to the chemical potential $\mu \sim -U/2$, and displays a $p$-wave–like angular dependence $\sim e^{-i\phi(\mathbf{k})}$. 
It gives rise to a single nodal line in the superconductor. Basically, along the direction of $\phi=\pm \frac{\pi}{2}$, the pairing projected on the band basis vanishes.
Both contributions are symmetry-allowed under the $C_{3z}$ rotation of TBG, 
with the first term more important around the $\Gamma$ point with small doping and the second term more important away from the $\Gamma$ point with larger doping. 
However, the second term is driven purely by interaction and therefore dominates over the first term when $M$ is small. Therefore, we expect a single nodal line in the superconductor.

\subsection{Effect of bandwidth $M$. }

In Fig.~\ref{fig:pandd} we show the STM spectrum for $w_0/w_1=0.8$, $\theta=1.06^\circ$, $J=6$ meV, $U=25$ meV, but with the active band width varying from $M=0$ to $M=8$ meV. 
A nodal SC structure is seen for both a very small $M$, and a very large $M$. 
The pairing structure is very different in the two regimes. 
For $M\approx 0$, the superconductor gap along each Fermi surface is $p_x$-wave like. 
For $M\approx 8$ meV, the superconductor gap has two nodal lines and is similar to $d_{x^2-y^2}$ pairing.

In the intermediate regime around $M\approx 3$ meV, 
the two hybridization terms compete with each other and the resulting STM spectrum shows no node. 
We note that even when the spectrum is not totally gapless, the gap is still momentum dependent, and it may still look like a $V$-shape in the STM spectrum.

\iffalse
\begin{equation}
\begin{aligned}
    &\epsilon_f(k) (|h|^2-|d_1|^2/2) s_+^\dagger s_+
    +\epsilon_f(k) (|h|^2-|d_2|^2/2) s_-^\dagger s_- \\
    &+ \sqrt{2}\epsilon_f(k) (hd_1 s^\dagger_{+K\uparrow}s^\dagger_{+K'\downarrow}
    + hd_2 s^\dagger_{-K\uparrow}s^\dagger_{-K'\downarrow})
\end{aligned}
\end{equation}

\begin{equation}
\begin{aligned}
    &(\delta_f(k) |h|^2-\delta^*_f(k) d_2^*d_1/2) s_+^\dagger s_-\\
    &+ \sqrt{1/2} ((\delta^*_f(k)hd_1 + \delta_f(k)hd_2) s^\dagger_{+K\uparrow}s^\dagger_{-K'\downarrow} + ...) + c.c.
\end{aligned}
\end{equation}

Written in terms of matrix:
\begin{equation}
\left[
    \begin{matrix}
        \epsilon-\mu & m_s & \Delta_{dig} & \Delta_{off}
    \end{matrix}
\right]
\end{equation}
\fi

\subsection{Comparison between the local pairing and the superconducting pairing.}

In Fig.~\ref{fig:deltak}, we show the momentum- and real-space pairing order parameters for both the mobile hole $s$ and netural spinon $\psi'$ fermions.
For the $s$ fermion, the pairing amplitude $\Delta_s(\mathbf{k})$ is concentrated near the Fermi surface, leading to a broader distribution in real space $\Delta_s(\mathbf{r}_i) = \frac{1}{\sqrt{N}}\sum_{\mathbf{k}} \Delta_s(\mathbf{k})e^{i\mathbf{k}\cdot \mathbf{r}_i}$. 
In contrast, the $\psi'$ fermion exhibits nearly uniform pairing in momentum space, with only weak $\mathbf{k}$ dependence near the Fermi surface. 
This corresponds to a localized pairing of $\psi'$ within a single AA site in real space. The superconductivity is mainly driven by pairing of the mobile hole $s$ fermions, and the cost of Hubbard $U$ is reduced because of its finite size in real space. Note that there is still a large on-site component, which is the artifact of the mean field calculation. In a more rigorous variational Monte Carlo (VMC) calculation using a Gutzwiller projected wavefunction to impose the constraint $n_{i;s}+n_{i;t}+\frac{1}{2}n_{i;\psi'} = 1$ exactly, this on-site contribution for pairing of $s$ would be removed.

\section{Gauge theory and confinement in the sFL phase}
\label{app:gauge}

In this section we discuss the gauge structure of our parton construction.  We note that in the superconductor (SC) and FL phase, all of the internal gauge fields are simply higgsed, so we do not need to worry about them. But the sFL phase needs some careful discussion.  Our parton construction is as follows:

\begin{equation}
    f_{i;\alpha}^{} = \sum_\beta s^\dagger_{i;\beta}\psi'_{i;\beta}\psi'_{i;\alpha}
    +\sum_{\beta\gamma} t^{}_{i;\alpha\beta\gamma} \psi'^\dagger_{i;\beta}\psi'^\dagger_{i;\gamma}~,
\end{equation}

Then there is an internal U(1) gauge field $a_\mu$ associated with the local gauge transformation: $\psi'_{i;\alpha} \rightarrow \psi'_{i;\alpha}e^{i \varphi_i}, s_{i;\alpha} \rightarrow s_{i;\alpha} e^{i 2 \varphi_i}, t_{i;\alpha \beta \gamma} \rightarrow t_{i;\alpha \beta \gamma} e^{i 2 \varphi_i}$.  We also introduce a probing field $A_\mu$ for the physical U(1) gauge field which generates the electric and magnetic field in physical world. In the end,  $s$ couples to $-A_\mu+2a_\mu$, $t$ couples to $A+2a_\mu$ and $\psi'$ couples to $a_\mu$. We interpret $\psi'$ as a neutral spinon, similar to the widely used Abrikosov fermion.

In the FL phase, we condense a slave boson $B \sim f^\dagger_\alpha \psi'_\alpha$, which couples to $-A_\mu+a_\mu$. After the condensation of  this Higgs boson $B$, we have $a_\mu=A_\mu$. Then the neutral spinon $\psi'$ is now charged under the physical U(1) field and has a finite spectral weight.

On the other hand, in the sFL phase, we have $\langle B \rangle =0$. Instead, we condense the spinon pairing term $\Delta \sim \psi_\alpha \psi_\beta$. For now let us ignore the flavor dependence. Note that $\Delta$ couples to $2a_\mu$ and its condensation only higgses the U(1) internal gauge field down to $Z_2$. Therefore, in our current ansatz of sFL phase, there is still a remaining $Z_2$ gauge field. We always have $f_i \sim s^\dagger \sim t$, so $s$ and $t$ now have finite overlap with the physical hole and electron operator.  If the remaining $Z_2$ gauge field does not confine, we should call the phase fractionalized Fermi liquid (FL*)\cite{senthil2003fractionalized} because there is a hidden topological order.   However, in 2+1d, $Z_2$ gauge field can be confined into a trivial and symmetric phase by condensing a vison ($\pi$ flux of the gauge field $a_\mu$) if the vison does not carry a non-trivial quantum number under symmetry.  For the usual case of an odd Mott insulator with one electron per site, vison always carries a momentum which forbids its condensation into a symmetric phase.  However, in our case, we have two electrons in the Mott insulator, and the vison does not have any quantum number. So both deconfined and confined phases are possible\cite{senthil2000z}, depending on energetics. At $\nu=-2$, they correspond to a $Z_2$ spin liquid and a trivial `rung singlet' phase. 

From our understanding, the local moments at $\nu=-2$ should just form a trivial and featureless `rung singlet' phase, so in this work we always assume that the Z$_2$ gauge field is confined. Such a confinement should persist near the sFL to SC transition.  In the sFL phase, $B$ couples to $-A_\mu$ and the $Z_2$ gauge field. Given the Z$_2$ gauge field is confined, at the sFL-SC transition, the slave boson $B$ is confined and the transition should be driven by the phase coherence of  a pair of $B$, which couples to $-2 A_\mu$ and can be identified as a physical Cooper pair.  Thus in the long distance limit, the sFL to SC transition should also be in the BKT universality class.  It would be interesting to study whether the slave boson $B$ can be probed in the short distance, such as in the core of the vortex of the superconductor.

\end{document}